\documentstyle[epsf]{livrev}
\begin{document}

\title{Computational Cosmology:
       from the Early Universe to the Large Scale Structure}

\author{Peter Anninos \\
        University of California \\
        Lawrence Livermore National Laboratory \\
        email:anninos1@llnl.gov \\ \\
       {\em (Accepted 15 February 2001)}  
}

\date{}
\maketitle


\begin{abstract}
In order to account for the observable Universe, 
any comprehensive theory or model of cosmology must draw
from many disciplines of physics, including gauge theories of
strong and weak interactions, the hydrodynamics and
microphysics of baryonic matter, 
electromagnetic fields, and spacetime curvature, for example.
Although it is difficult
to incorporate all these physical elements into a 
single complete model of our Universe,
advances in computing methods and technologies
have contributed significantly towards our
understanding of cosmological models, the Universe,
and astrophysical processes within them.
A sample of numerical calculations (and numerical
methods) applied to specific
issues in cosmology are reviewed in this article: from
the Big Bang singularity dynamics to the fundamental
interactions of gravitational waves; from the 
quark-hadron phase transition to the large scale
structure of the Universe. 
The emphasis, although not exclusively, is on those
calculations designed to test different models
of cosmology against the observed Universe.
\end{abstract}


\newpage



\section{Introduction}
\label{sec:intro}

Numerical investigations of cosmological spacetimes can be
categorized into two broad classes of calculations, distinguished
by their computational (or even philosophical) goals:
1) geometrical and mathematical principles of 
cosmological models, and
2) physical and astrophysical cosmology.
In the former, the emphasis is on the geometric
framework in which astrophysical processes
occur, namely the cosmological expansion, shear, and
singularities of the many models allowed by the theory of general
relativity. In the latter, the emphasis is on the
cosmological and
astrophysical processes in the real or observable Universe, and
the quest to determine the model which best describes our Universe.
The former is pure in the sense that it concerns the fundamental
nonlinear behavior of the Einstein equations and the
gravitational field.
The latter is more complex as it
addresses the composition, organization,
and dynamics of the Universe from the small scales 
(fundamental particles and elements)
to the large (galaxies and clusters of galaxies).
However the distinction is not always so clear, and geometric
effects in the spacetime curvature can have significant
consequences for the evolution and observation of matter
distributions.

Any comprehensive model of cosmology must
therefore include nonlinear interactions
between different matter sources and spacetime curvature.
A realistic model of the Universe must also cover large
dynamical spatial and temporal scales, extreme temperature and
density distributions, and highly dynamic atomic
and molecular matter compositions.
In addition, due to all the varied physical processes of
cosmological significance, one must draw from many
disciplines of physics to model
curvature anisotropies, gravitational waves,
electromagnetic fields, nucleosynthesis,
particle physics, hydrodynamic fluids, etc.
These phenomena are described in terms of coupled nonlinear
partial differential equations and must be solved numerically
for general inhomogeneous spacetimes.
The situation appears extremely complex, even with current
technological and computational advances.
As a result, the codes and numerical methods
that have been developed to date are designed
to investigate very specific problems with
either idealized symmetries or simplifying assumptions regarding
the metric behavior, the matter distribution/composition
or the interactions among the matter types and spacetime curvature.

It is the purpose of this article to review published
numerical cosmological calculations addressing problems 
from the very early Universe to the present;
from the purely geometrical dynamics of the initial singularity
to the large scale structure of the Universe.
There are three major sections:
\S~\ref{sec:background} where a brief overview is presented
of various defining events occurring throughout the history of our Universe
and in the context of the standard model;
\S~\ref{sec:relcosmology} where summaries
of early Universe and relativistic
cosmological calculations are presented;
and \S~\ref{sec:physcosmology} which focuses
on structure formation in the post-recombination epoch
and on testing cosmological models against observations.
Following a few conclusion statements in \S~\ref{sec:conclusion},
an appendix \S~\ref{sec:appendix} discusses the basic
Einstein equations, kinematic considerations, 
matter source equations with curvature, and the equations of
perturbative physical cosmology on background isotropic
models. References to numerical methods are also
supplied and reviewed for each case.

\newpage


\section{Background}
\label{sec:background}


\subsection{A brief chronology}
\label{subsec:timeline}

With current observational constraints, the physical state of our
Universe, as understood in the context of the standard 
or Friedmann-Lema\^\i tre-Robertson-Walker (FLRW) model,
can be crudely extrapolated back to $\sim 10^{-34}$ seconds after the
Big Bang, before which the classical description of general relativity
is expected to give way to a quantum theory of gravity.
At the earliest times, the Universe was a plasma of
relativistic particles consisting of quarks, leptons,
gauge bosons, and Higgs bosons represented by
scalar fields with interaction and symmetry regulating potentials.
It is believed that several spontaneous
symmetry breaking (SSB) phase transitions occured in
the early Universe as it expanded and cooled,
including the grand unification
transition (GUT) at $\sim 10^{-34}$ seconds
after the Big Bang in which the strong nuclear force split off
from the weak and electromagnetic forces (this also marks an era
of inflationary expansion and the
origin of matter-antimatter asymmetry through baryon,
charge conjugation, and charge + parity violating interactions
and nonequilibrium effects); the electroweak (EW)
SSB transition at $\sim 10^{-11}$~s when the weak nuclear
force split from the electromagnetic force; and the chiral or quantum
chromodynamic (QCD)
symmetry breaking transition at $\sim 10^{-5}$~s during which
quarks condensed into hadrons. The most stable hadrons (baryons,
or protons and neutrons comprised of three quarks) survived the
subsequent period of baryon-antibaryon annihilations, which
continued until the Universe cooled to the point at which 
new baryon-antibaryon pairs could no longer be produced. This resulted
in a large number of photons and relatively few surviving baryons.
A period of primordial nucleosynthesis followed from
$\sim 10^{-2}$ to $\sim 10^2$~s during which light element
abundances were synthesized to form 24\% helium with trace amounts of
deuterium, tritium, helium-3, and lithium.

By $\sim 10^{11}$~s, the matter density became equal to the
radiation density as the Universe continued to 
expand, identifying the start of
the current matter-dominated
era and the beginning of structure formation.
Later, at $\sim 10^{13}$~s ($3\times10^5$ years), 
the free ions and electrons combined to form atoms,
effectively decoupling the matter from the radiation field as
the Universe cooled. This decoupling
or post-recombination epoch
marks the surface of last scattering and the boundary
of the observable (via photons) Universe.
Assuming a hierarchical Cold Dark Matter (CDM) structure formation
scenario, the subsequent development 
of our Universe is characterized by the
growth of structures with increasing size. For example,
the first stars are likely to have formed at $t\sim10^8$ years
from molecular gas clouds when the Jeans mass of the
background baryonic fluid was approximately $10^4~M_\odot$,
as indicated in Figure~\ref{fig:reheating}.
This epoch of pop~III star generation is followed by the formation
of galaxies at $t\sim10^9$ years and subsequently galaxy clusters.
Though somewhat controversial, estimates of the current
age of our Universe range from 10 to 20 Gyrs, with
a present-day linear structure scale radius of about
$8 \, h^{-1}$ Megaparsecs, where $h$ is the Hubble
parameter (compared to 2--3 Megaparsecs
typical for the virial radius of rich galaxy clusters).

\begin{figure}[htbp]
  \def\epsfsize#1#2{0.5#1}
  \caption{\it Schematic depicting the general sequence of events in
    the post-recombination Universe. The solid and dotted lines
    potentially track the Jeans mass of the average baryonic gas
    component from the recombination epoch at $z\sim10^3$ to the
    current time. A residual ionization fraction of
    $n_{\rm H+}/n_{\rm H} \sim 10^{-4}$ following recombination allows
    for Compton interactions with photons to $z\sim200$, during which
    the Jeans mass remains constant at $10^5 M_{\odot}$. The Jeans
    mass then decreases as the Universe expands adiabatically until
    the first collapsed structures form sufficient amounts of hydrogen
    molecules to trigger a cooling instability and produce pop\protect~III
    stars at $z\sim20$. Star formation activity can then reheat  the
    Universe and raise the mean Jeans mass to above
    $10^8 M_\odot$. This reheating could affect the subsequent
    development of structures such as galaxies and the observed
    Lyman-alpha clouds.}
  \label{fig:reheating}
\end{figure}


\subsection{Successes of the standard model}
\label{subsec:standard}

The isotropic and homogeneous FLRW cosmological model
has been so successful in describing the observable Universe
that it is commonly referred to as the ``standard model''.
Furthermore, and to its credit, the model is relatively simple
so that it allows for calculations and predictions
to be made of the very early Universe, including
primordial nucleosynthesis at $10^{-2}$ seconds
after the Big Bang, and even
particle interactions approaching the Planck
scale at $10^{-43}$ seconds.
At present, observational support for the standard model includes:

\begin{itemize}
\item {\it the expansion of the Universe}
  as verified by the redshifts
  in galaxy spectra and quantified by measurements of
  the Hubble constant $H_0=100 h {\rm\ km\ s}^{-1}$ ${\rm Mpc}^{-1}{\rm\ }$; 
\item {\it the deceleration parameter}
  observed in
  distant galaxy spectra (although uncertainties about
  galactic evolution, intrinsic luminosities,
  and standard candles prevent an accurate estimate);
\item {\it the large scale isotropy and homogeneity of the Universe}
  based on temperature anisotropy measurements of the
  microwave background radiation and
  peculiar velocity fields of galaxies
  (although the light distribution from bright galaxies is
  somewhat contradictory);
\item {\it the age of the Universe}
  which yields roughly consistent estimates between the look-back
  time to the Big Bang in the FLRW model and observed data such
  as the oldest stars, radioactive elements, and cooling of
  white dwarf stars;
\item {\it the cosmic microwave background radiation}
  suggests that the Universe began from a hot Big Bang
  and the data is consistent with a mostly isotropic model and a black
  body at temperature 2.7~K;
\item {\it the abundance of light elements}
  such as $^2$H, $^3$He, $^4$He and $^7$Li, as predicted from
  the FLRW model, is consistent with observations and
  provides a bound on the baryon density and baryon-to-photon
  ratio;
\item {\it the present mass density},
  as determined from measurements of
  luminous matter and galactic rotation curves,
  can be accounted for by the FLRW model with a single density
  parameter ($\Omega_0$) to specify the metric topology;
\item {\it the distribution of galaxies and larger scale structures}
  can be reproduced by numerical simulations
  in the context of inhomogeneous perturbations of the FLRW models.
\end{itemize}

Because of these successes, most work in the field 
of physical cosmology (see \S~\ref{sec:physcosmology})
has utilized the standard model as the background spacetime
in which the large scale structure evolves, with the
ambition to further constrain parameters 
and structure formation scenarios
through numerical simulations.
The reader is referred to~\cite{KT90}
for a more in-depth review of the standard model,
and to~\cite{OS95, TZH00}
for a summary of observed cosmological
parameter constraints and best fit ``concordance'' models.

\newpage


\section{Relativistic Cosmology}
\label{sec:relcosmology}

This section is organized to track the chronological
events in the history of the early or relativistic Universe, 
focusing mainly on four defining moments:
1)~the Big Bang singularity and the dynamics of the
very early Universe; 2)~inflation and its generic nature;
3)~QCD phase transitions; and
4)~primordial nucleosynthesis
and the freeze-out of the light elements.
The late or post-recombination epoch
is reserved to a separate section \S~\ref{sec:physcosmology}.


\subsection{Singularities}
\label{sec:singularity}


\subsubsection{Mixmaster dynamics}
\label{sec:mixmaster}

Belinsky, Lifshitz and Khalatnikov (BLK)~\cite{BLK70, BLK71}
and Misner~\cite{Misner69} discovered
that the Einstein equations in the vacuum homogeneous Bianchi
type IX (or Mixmaster) cosmology exhibit
complex behavior and are sensitive
to initial conditions as the Big Bang singularity is approached.
In particular, the solutions near the singularity are
described qualitatively by a discrete
map~\cite{Barrow81, BLK70}
representing different sequences of Kasner spacetimes
\begin{equation}
  ds^2 = -dt^2 + {t}^{2p_1} dx^2 + {t}^{2p_2} dy^2 + {t}^{2p_3} dz^2,
  \label{eqn:kasner}
\end{equation}
with time changing exponents $p_i$, but otherwise constrained by
$p_1+p_2+p_3=p_1^2+p_2^2+p_3^2=1$.
Because this discrete mapping of Kasner epochs is chaotic,
the Mixmaster dynamics is presumed to be chaotic as well.

\begin{figure}[htbp]
  \def\epsfsize#1#2{0.5#1}
  \caption{\it Contour plot of the Bianchi type IX potential $V$, where
    $\beta_{\pm}$ are the anisotropy canonical coordinates. Seven level
    surfaces are shown at equally spaced decades ranging from $10^{-1}$
    to $10^5$. For large isocontours ($V>1$), the potential is open  and
    exhibits a strong triangular symmetry with three narrow channels
    extending to spatial infinity. For $V<1$, the potential closes and
    is approximately circular for $\beta_{\pm} \ll 1$.}
  \label{fig:mixmaster}
\end{figure}

Mixmaster behavior can be studied in the context of Hamiltonian
dynamics, with a Hamiltonian~\cite{MTW73}
\begin{equation}
  2{\cal H} = -p_\Omega^2 + p_+^2 + p_-^2 + e^{4\alpha} (V-1),
  \label{eqn:mixmaster1}
\end{equation}
and a semi-bounded potential arising from the spatial scalar
curvature (whose level curves are plotted in
Figure~\ref{fig:mixmaster})
\begin{equation}
  V = 1 + \frac{1}{3}e^{-8\beta_+} +
  \frac{2}{3}e^{4\beta_+}\left[\cosh(4\sqrt{3}\beta_-) -1\right]
  -\frac{4}{3}e^{-2\beta_+}\cosh(2\sqrt{3}\beta_-),
  \label{eqn:mixmaster2}
\end{equation}
where $e^\alpha$ and $\beta_{\pm}$ are the scale factor and
anisotropies, and $p_\alpha$ and $p_{\pm}$ are the corresponding
conjugate variables. A solution of this Hamiltonian system
is an infinite sequence of Kasner epochs with
parameters that change when the phase space trajectories
bounce off the potential walls, which become exponentially
steep as the system evolves towards the singularity.
Several numerical calculations of the dynamical equations
arising from~(\ref{eqn:mixmaster1}) and~(\ref{eqn:mixmaster2})
have indicated that the Liapunov exponents of the system vanish,
in apparent contradiction with the discrete maps~\cite{BBE90, HBWS91},
and putting into question the characterization of Mixmaster dynamics
as chaotic. However, it has since been
shown that the usual definition of the
Liapunov exponents is ambiguous in this case
as it is not invariant under time reparametrizations, and that
with a different time variable one obtains positive
exponents~\cite{Berger91, FFM91}. Also,
coordinate independent methods using fractal basin boundaries
to map basins of attraction in the space of initial
conditions indicates Mixmaster spacetimes to be chaotic~\cite{CL97}.

Although BLK conjectured that local Mixmaster oscillations
might be the generic behavior for singularities in  more
general classes of spacetimes~\cite{BLK71},
it is only recently that this conjecture has begun to
be supported by numerical evidence (see Section~\ref{sec:avtd}
and~\cite{Berger98}).


\subsubsection{AVTD vs.\ BLK oscillatory behavior}
\label{sec:avtd}

As noted in \S~\ref{sec:mixmaster},
an interesting and important issue in classical cosmology is
whether or not the generic Big Bang
singularity is locally of a Mixmaster 
or BLK type, with complex
oscillatory behavior as the singularity is approached.
Most of the Bianchi models,
including the Kasner solutions~(\ref{eqn:kasner}),
are characterized by either
open or no potentials in the Hamiltonian framework~\cite{RS75},
and exhibit essentially monotonic or
Asymptotically Velocity Term Dominated (AVTD) behavior.

Considering inhomogeneous spacetimes, Isenberg and 
Moncrief~\cite{IM90} proved that the singularity
in the polarized Gowdy model is AVTD type, as are more
general polarized $T^2$ symmetric cosmologies~\cite{BCIM97}.
Early numerical studies using symplectic
methods have confirmed these conjectures
and found no evidence of BLK oscillations, even in
$T^3\times R$ spacetimes with $U(1)$ symmetry~\cite{Berger95}
(although there were concerns about the solutions due to
difficulties in resolving steep spatial gradients
near the singularity~\cite{Berger95}), which were addressed later by
Hern and Stewart~\cite{HS98} for the Gowdy $T^3$ models).
However, Weaver et al.~\cite{WIB98} have established
the first evidence through numerical simulations that
Mixmaster dynamics can occur in (at least a restricted class of)
inhomogeneous spacetimes which generalize the Bianchi
type VI$_0$ with a magnetic field and two-torus symmetry.
More recently, Berger and Moncrief~\cite{BM98, BM00} have
shown $U(1)$ symmetric vacuum cosmologies to exhibit local
Mixmaster dynamics, which tends to support the BLK conjecture.
Despite numerical difficulties in resolving steep
gradients (which they managed by enforcing the Hamiltonian
constraint and spatially averaging the solutions), 
Berger and Moncrief have confirmed their findings
under increased spatial resolution and changes in initial
data.


\subsection{Inflation}
\label{sec:inflation}

The inflation paradigm is frequently invoked to explain the flatness
($\Omega_0\approx 1$ in the context of the FLRW model) and nearly
isotropic nature of the Universe at large scales, attributing them to
an era of exponential expansion at about $10^{-34} {\rm\ s}$ after the
Big Bang. This expansion acts as a strong dampening mechanism to
random curvature or density fluctuations, and may be a generic
attractor in the space of initial conditions. An essential component
needed to trigger this inflationary phase is a scalar or inflaton
field $\phi$ representing spin zero particles. The vacuum energy of
the field acts as an effective cosmological constant that regulates
GUT symmetry breaking, particle creation, and the reheating of the
Universe through an interaction potential $V(\phi)$ derived from the
form of symmetry breaking that occurs as the temperature of the
Universe decreases. Early analytic studies focused on simplified
models, treating the interaction potential as flat near its local
maximum where the field does not evolve significantly and where the
formal analogy to an effective cosmological constant approximation is
more precise. However, to properly account for the complexity of
inflaton fields, the full dynamical equations (see
\S~\ref{sec:scalar}) must be considered together with consistent
curvature, matter and other scalar field couplings.  Also, many
different theories of inflation and vacuum potentials have been
proposed (see, for example, a recent review by Lyth and
Riotto~\cite{LR99} and an introductory article by
Liddle~\cite{Liddle99}), and it is not likely that simplified models
can elucidate the full nonlinear complexity of scalar fields (see
\S~\ref{sec:scalarfielddynamics}) nor the generic nature of inflation.

In order to study whether inflation can occur for arbitrary
anisotropic and inhomogeneous data, many numerical simulations
have been carried out with different symmetries,
matter types and perturbations. A sample of such
calculations is described in the following paragraphs.


\subsubsection{Plane symmetry}
\label{sec:inflation_plane}

Kurki-Suonio et al.~\cite{KCMW87} extended the planar
cosmology code
of Centrella and Wilson~\cite{CW83, CW84} (see \S~\ref{sec:gwaves})
to include a scalar field and
simulate the onset of inflation in the early Universe
with an inhomogeneous Higgs field and a perfect
fluid with a radiation equation of state $p=\rho/3$, where
$p$ is the pressure and $\rho$ is the energy density.
Their results suggest that whether inflation occurs or not can be
sensitive to the shape of the potential $\phi$. In particular,
if the shape is flat enough and satisfies the slow-roll conditions
(essentially upper bounds on $\partial V/\partial\phi$ and
$\partial^2 V/\partial\phi^2$~\cite{KT90} 
near the false vacuum $\phi \sim 0$),
even large initial fluctuations
of the Higgs field do not prevent inflation.
They considered two different forms of the potential: a
Coleman-Weinberg type with interaction strength $\lambda$
\begin{equation}
  V(\phi)=\lambda\phi^4\left[\ln\left(\frac{\phi^2}{\sigma^2}\right)
    -\frac{1}{2}\right]+\frac{\lambda\sigma^4}{2}
\end{equation}
which is very flat close
to the false vacuum and does inflate; and a rounder
``$\phi^4$'' type
\begin{equation}
  V(\phi)=\lambda(\phi^2-\sigma^2)^2
\end{equation}
which, for their parameter combinations, does not.


\subsubsection{Spherical symmetry}
\label{sec:spherical}

Goldwirth and Piran~\cite{GP90} studied
the onset of inflation with inhomogeneous initial
conditions for closed, spherically symmetric spacetimes
containing a massive scalar field and radiation field sources
(described by a massless scalar field). 
In all the cases they considered, 
the radiation field damps quickly and only
an inhomogeneous massive scalar field remains
to inflate the Universe. They find that
small inhomogeneities tend to reduce the amount of inflation
and large initial inhomogeneities can even suppress the onset of
inflation. Their calculations indicate that
the scalar field must have ``suitable'' initial values
over a domain of several horizon lengths in order 
for inflation to begin.


\subsubsection{Bianchi V}
\label{sec:inflation_bianchiV}

Anninos et al.~\cite{AMRR91} investigated the simplest Bianchi
model (type V) background that admits velocities or tilt in order
to address the question of how the Universe can choose a uniform
reference frame at the exit from inflation, since the
de Sitter metric does not have a preferred frame. They find that
inflation does not isotropize the Universe in the short wavelength limit.
However, if inflation persists, the wave behavior eventually freezes
in and all velocities go to zero at least as rapidly as
$\tanh\beta \sim R^{-1}$, where $\beta$ is the relativistic tilt angle
(a measure of velocity), and $R$ is a typical scale 
associated with the radius of the Universe.
Their results indicate
that the velocities eventually go to zero as inflation carries 
all spatial variations outside the horizon, and that the answer
to the posed question is that memory is retained and the
Universe is never really de Sitter.


\subsubsection{Gravity waves + cosmological constant}
\label{sec:inflation_gwaves}

In addition to the inflaton field, one can consider other
sources of inhomogeneity, such as gravitational waves.
Although linear waves in de Sitter space will
decay exponentially and disappear, it is unclear what
will happen if strong waves exist. 
Shinkai \& Maeda~\cite{SM93} investigated
the cosmic no-hair conjecture
with gravitational waves and a cosmological constant ($\Lambda$)
in 1D plane symmetric vacuum spacetimes, setting 
up Gaussian pulse wave data with
amplitudes $0.02 \Lambda \le \mbox{max}(\sqrt{I}) \le 80 \Lambda$
and widths $0.08 \, l_{\rm H} \le l \le 2.5 \, l_{\rm H}$, where
$I$ is the invariant constructed from the 3-Riemann tensor
and $l_{\rm H} = \sqrt{3/\Lambda}$ is the horizon scale.
They also considered colliding plane waves with
amplitudes $40 \Lambda \le \mbox{max}(\sqrt{I}) \le 125 \Lambda$
and widths $0.08 \, l_{\rm H} \le l \le 0.1 \, l_{\rm H}$.
They find that for any large amplitude
or small width inhomogeneity in their parameter sets,
the nonlinearity
of gravity has little effect and the spacetime always
evolves towards de Sitter.


\subsubsection{3D inhomogeneous spacetimes}
\label{sec:inflation_3d}

The previous paragraphs discussed results from highly
symmetric spacetimes, but the possibility of
inflation remains to be established for more
general inhomogeneous and nonperturbative data.
In an effort to address this issue,
Kurki-Suonio et al.~\cite{KLM93} investigated
fully three-dimensional inhomogeneous
spacetimes with a chaotic inflationary potential
$V(\phi) =\lambda\phi^4/4$. They
considered basically two types of runs: small and large scale.
In the small scale run, the grid
length was initially set equal to the Hubble length so the
inhomogeneities are well inside the horizon and the dynamical
time scale is shorter than the expansion or Hubble time.
As a result, the perturbations oscillate and damp, while
the field evolves and the spacetime inflates.
In the large scale run, the inhomogeneities are outside
the horizon and they do not oscillate. They maintain their
shape without damping and,
because larger values of $\phi$ lead to faster
expansion, the inhomogeneity in the expansion becomes
steeper in time since the regions of large $\phi$ and
high inflation stay correlated.
Both runs have sufficient inflation to solve the
flatness problem.


\subsection{Chaotic scalar field dynamics}
\label{sec:scalarfielddynamics}

Many studies of cosmological models generally reduce complex physical
systems to simplified or even analytically
integrable systems. In sufficiently simple models the 
dynamical behavior (or fate) of the Universe
can be predicted from a given set of initial conditions.
However, the Universe is composed of many different nonlinear
interacting fields including the inflaton field which can
exhibit complex chaotic behavior. For example, 
Cornish and Levin~\cite{CL96} consider a homogenous
isotropic closed FLRW model with various conformal and minimally
coupled scalar fields (see \S~\ref{sec:scalar}). They find
that even these relatively simple models exhibit chaotic transients
in their early pre-inflationary evolution. This behavior in exiting
the Planck era is characterized by fractal basins of attraction,
with attractor states being to (1) inflate forever, (2) inflate
over a short period of time then collapse, or (3) collapse
without inflating.
Monerat et al.~\cite{MOS98} investigated
the dynamics of the pre-inflationary phase of the Universe
and its exit to inflation in a closed FLRW model with
radiation and a minimally coupled scalar field. They
observe complex behavior associated with saddle-type critical
points in phase space that give rise to 
deSitter attractors with multiple chaotic exits to inflation
that depend on the structure of the scalar field potential.
These results suggest that distinctions between exits to
inflation may be manifested in the process
of reheating and as a selected spectrum of
inhomogeneous perturbations influenced by resonance mechanisms
in curvature oscillations. This could possibly lead
to fractal patterns in the density spectrum, gravitational waves,
CMBR field, or galaxy distribution that depend
on specific details including the number of fields,
the nature of the fields, and their interaction potentials.

Chaotic behavior can also be found in more general applications
of scalar field dynamics.
Anninos et al.~\cite{AOM91} investigated the nonlinear
behavior of colliding kink-antikink solitons or domain walls
described by a single real scalar field with self-interaction
potential $\lambda(\phi^2 - 1)^2$. Domain walls can form
as topological defects during the spontaneous symmetry breaking
period associated with phase transitions, and can seed
density fluctuations in the large scale structure.
For collisional time scales much smaller 
than the cosmological expansion, they find that
whether a kink-antikink collision results
in a bound state or a two-soliton solution depends on
a fractal structure observed in the impact velocity parameter space.
The fractal structure arises from a resonance condition
associated with energy exchanges between translational modes
and internal shape-mode oscillations. The impact parameter
space is a complex self-similar fractal composed of sequences
of different $n$-bounce (the number of bounces or oscillations
in the transient semi-coherent state) reflection windows 
separated by regions of oscillating bion states 
(see Figure~\ref{fig:fractal}).
They compute the Lyapunov exponents for parameters in which
a bound state forms to
demonstrate the chaotic nature of the bion oscillations.

\begin{figure}[htbp]
  \def\epsfsize#1#2{0.5#1}
  \caption{\it Fractal structure of the transition between reflected
    and captured states for colliding kink-antikink solitons in the
    parameter space of impact velocity for a $\lambda(\phi^2-1)^2$
    scalar field potential. The top image (a) shows the 2-bounce
    windows in dark with the rightmost region ($v/c > 0.25$)
    representing the single-bounce regime above which no captured
    state exists, and the leftmost white region ($v/c < 0.19$)
    representing the captured state below which no reflection windows
    exist. Between these two marker velocities, there are 2-bounce
    reflection states of decreasing widths separated by regions of
    bion formation. Zooming in on the domain outlined by the dashed
    box, a self-similar structure is apparent in the middle image (b),
    where now the dark regions represent 3-bounce windows of
    decreasing widths. Zooming in once again on the boundaries of
    these 3-bounce windows, a similar structure is found as shown in
    the bottom image (c) but with 4-bounce reflection windows. This
    pattern of self-similarity with $n$-bounce windows is observed at
    all scales investigated numerically.}
  \label{fig:fractal}
\end{figure}


\subsection{Quark-hadron phase transition}
\label{sec:quark}

The standard picture of cosmology assumes that a phase transition 
(associated with chiral symmetry breaking after the
electroweak transition) occurred
at approximately $10^{-5}$ seconds after the Big Bang to convert
a plasma of free quarks and gluons into hadrons. 
Although this transition can be of significant cosmological
importance, it is not known with certainty whether
it is of first order or higher, and what the
astrophysical consequences might be
on the subsequent state of the Universe. For example, the
transition may give rise to significant baryon number inhomogeneities
which can influence the outcome of primordial nucleosynthesis 
as evidenced in the distribution and averaged light element abundances.
The QCD transition and baryon inhomogeneities 
may also play a significant and potentially
observable role in the generation of primordial magnetic fields.

Rezolla et al.~\cite{RMP95}
considered a first order phase transition and
the nucleation of hadronic bubbles in
a supercooled quark-gluon plasma,
solving the relativistic Lagrangian equations for
disconnected and evaporating quark regions during the
final stages of the phase transition.
They numerically investigated a single
isolated quark drop with an initial radius large enough
so that surface effects can be neglected.
The droplet evolves as a self-similar solution
until it evaporates to a sufficiently small radius that
surface effects break the similarity solution and
increase the evaporation rate.
Their simulations indicate that, in neglecting
long-range energy and momentum transfer
(by electromagnetically interacting particles)
and assuming that the baryon number is transported
with the hydrodynamical flux, the baryon number
concentration is similar to what is predicted by
chemical equilibrium calculations.

Kurki-Suonio and Laine~\cite{KL95}
studied the growth of bubbles and the decay of droplets
using a spherically symmetric code that
accounts for a phenomenological model
of the microscopic entropy generated at the
phase transition surface. 
Incorporating the small scale effects of the finite wall width
and surface tension, but neglecting entropy and baryon flow
through the droplet wall, they 
demonstrate the dynamics of nucleated 
bubble growth and quark droplet decay.
They also find that evaporating droplets do not leave behind a
global rarefaction wave to dissipate any previously generated
baryon number inhomogeneity.


\subsection{Nucleosynthesis}
\label{sec:nucleo}

Observations of the light elements produced during
Big Bang nucleosynthesis following the quark/hadron 
phase transition
(roughly $10^{-2}$--$10^{2}$
seconds after the Big Bang) are in good agreement
with the standard
model of our Universe (see \S~\ref{subsec:standard}).
However, it is interesting to investigate other
more general models to assert the role of shear
and curvature on the nucleosynthesis process.

Rothman and Matzner~\cite{RM84} 
considered primordial nucleosynthesis in an\-iso\-tropic cosmologies,
solving the strong reaction equations leading to $^4$He.
They find that the concentration of $^4$He increases
with increasing shear due to time scale effects and the
competition between dissipation and enhanced
reaction rates from photon heating and neutrino
blue shifts. Their results have been used to
place a limit on anisotropy at the epoch of nucleosynthesis.
Kurki-Suonio and Matzner~\cite{KM85} extended this work
to include 30 strong 2-particle reactions involving nuclei
with mass numbers $A\le 7$, and to
demonstrate the effects of anisotropy on the 
cosmologically significant isotopes $^2$H, $^3$He,
$^4$He and $^7$Li as a function of the baryon to photon ratio.
They conclude that the effect of anisotropy
on $^2$H and $^3$He is not significant, and the abundances
of $^4$He and $^7$Li increase with anisotropy in accord
with~\cite{RM84}.

Furthermore, it is possible that neutron diffusion,
the process whereby neutrons diffuse out from regions of very
high baryon density just before nucleosynthesis,
can affect the neutron to proton ratio in such a way as to
enhance deuterium and reduce $^4$He compared to a
homogeneous model. However, plane symmetric, general relativistic
simulations with neutron diffusion~\cite{KMCRW88} show
that the neutrons diffuse back into the high density regions
once nucleosynthesis begins there -- thereby wiping out the effect.
As a result, although inhomogeneities influence the element
abundances, they do so at a much smaller degree then
previously speculated. The 
numerical simulations also demonstrate that, because
of the back diffusion, a cosmological
model with a critical baryon density
cannot be made consistent with helium and deuterium
observations, even with substantial baryon inhomogeneities
and the anticipated neutron diffusion effect.


\subsection{Plane symmetric gravitational waves}
\label{sec:gwaves}

Gravitational waves are an inevitable
product of the Einstein equations, and one can expect
a wide spectrum of wave signals propagating throughout
our Universe due to shear anisotropies, primordial
metric and matter fluctuations, collapsing matter
structures, ringing black holes, and
colliding neutron stars, for example. The discussion
here is restricted to the pure vacuum field
dynamics and the fundamental nonlinear behavior
of gravitational waves in numerically
generated cosmological spacetimes.

Centrella and Matzner~\cite{CM79, CM82} studied
a class of plane symmetric cosmologies representing
gravitational inhomogeneities in the form of shocks or
discontinuities separating two vacuum expanding
Kasner cosmologies~(\ref{eqn:kasner}). 
By a suitable choice of parameters,
the constraint equations can be
satisfied at the initial time with an Euclidean 3-surface and
an algebraic matching of parameters
across the different Kasner regions
that gives rise to a discontinuous extrinsic curvature tensor.
They performed both numerical calculations and analytical
estimates using a Green's function analysis to establish
and verify (despite the numerical
difficulties in evolving discontinuous
data) certain aspects of the solutions, including
gravitational wave interactions, the formation of tails, 
and the singularity behavior of colliding waves in
expanding vacuum cosmologies.

Shortly thereafter, Centrella and Wilson~\cite{CW83, CW84}
developed a polarized plane symmetric code for cosmology,
adding also hydrodynamic sources with artificial viscosity
methods for shock capturing and Barton's method for 
monotonic transport~\cite{Wilson79}.
The evolutions are fully constrained (solving both the
momentum and Hamiltonian constraints at each time step) and use the
mean curvature slicing condition.
This work was subsequently extended by 
Anninos et al.~\cite{ACM89, ACM91a, Anninos98},
implementing more robust numerical methods, an improved
parametric treatment of the initial value problem,
and generic unpolarized metrics.

In applications of these codes,
Centrella~\cite{Centrella86} investigated nonlinear gravity waves
in Minkowski space and compared the full numerical 
solutions against a first
order perturbation solution to benchmark certain numerical
issues such as numerical damping and dispersion. A second
order perturbation analysis was used to model the
transition into the nonlinear regime.
Anninos et al.~\cite{ACM91b} considered small and large
perturbations in the two degenerate Kasner models:
$p_1=p_3=0$ or $2/3$, and $p_2=1$ or $-1/3$
respectively, where $p_i$ are parameters
in the Kasner metric~(\ref{eqn:kasner}).
Carrying out a second order perturbation expansion
and computing the Newman-Penrose (NP) scalars, Riemann
invariants and Bel-Robinson vector, they demonstrated,
for their particular class of spacetimes, that the
nonlinear behavior is in the Coulomb (or background) part
represented by the leading order term in the NP scalar $\Psi_2$,
and not in the gravitational wave component.
For standing-wave perturbations,
the dominant second order effects in their variables
are an enhanced monotonic increase in the background expansion rate,
and the generation of oscillatory behavior in the background
spacetime with frequencies equal to the harmonics
of the first order standing-wave solution.
Expanding their investigations of the
Coulomb nonlinearity, Anninos and McKinney~\cite{AM99}
used a gauge invariant perturbation formalism to
construct constrained initial data for 
general relativistic cosmological sheets formed
from the gravitational collapse of an ideal gas in a
critically closed FLRW ``background'' model.
Results are compared to the Newtonian Zel'dovich~\cite{Zeldovich70} solution over a range of
field strengths and flows. Also, the growth rates
of nonlinear modes (in both the gas density and
Riemann curvature invariants), their effect in 
the back-reaction to modify the cosmological
scale factor, and their role in generating
CMB anisotropies are discussed.


\subsection{Regge calculus model}
\label{sec:regge}

A unique approach to numerical cosmology (and numerical
relativity in general) is the method of Regge Calculus 
in which spacetime is represented as a complex
of 4-dimensional, geometrically flat simplices.
The principles of Einstein's theory are applied
directly to the simplicial geometry to form the curvature,
action, and field equations, in contrast to the
finite difference approach where the continuum field
equations are differenced on a discrete mesh.

A 3-dimensional code implementing Regge Calculus
techniques was developed recently by
Gentle and Miller~\cite{GM98} and applied to the Kasner
cosmological model.
They also describe a procedure to solve
the constraint equations for time
asymmetric initial data on two spacelike hypersurfaces
constructed from tetrahedra, since full
4-dimensional regions or lattices are used.
The new method is analogous to York's
procedure (see~\cite{York79} and \S~\ref{sec:initial})
where the conformal metric,
trace of the extrinsic curvature, and momentum 
variables are all freely specifiable.
These early results are promising in that they have reproduced
the continuum Kasner solution, achieved second order 
convergence, and sustained numerical stability.
Also, Barnett et al.~\cite{BGMSTW97} discuss an implicit
evolution scheme that allows
local (vertex) calculations for efficient parallelism.
However, the Regge Calculus approach remains to be developed
and applied to more general spacetimes with complex
topologies, extended degrees of freedom, and general
source terms.

\newpage


\section{Physical Cosmology}
\label{sec:physcosmology}

The phrase ``physical cosmology'' is generally associated 
with the large (galaxy and cluster) 
scale structure of the post-recombination epoch
where gravitational effects are modeled approximately
by Newtonian physics on a uniformly expanding,
matter dominated FLRW background.
A discussion of the large scale
structure is included in this review since any viable model 
of our Universe which allows a regime where strongly general 
relativistic effects are important must
match onto the weakly relativistic
(or Newtonian) regime. Also, since certain aspects
of this regime are directly observable, one can hope to
constrain or rule out various cosmological models and/or parameters,
including the density ($\Omega_0$), 
Hubble ($H_0=100 \, h {\rm\ km\ s}^{-1}{\rm\ Mpc}^{-1}$),
and cosmological ($\Lambda$) constants.

Due to the vast body of literature on numerical simulations
of the post-recombination epoch,
it is possible to mention only a very small fraction
of all the published papers. Hence, the following
summary is limited to cover just a few aspects of computational
physical cosmology, and in particular
those that can potentially be used to
discriminate between cosmological model parameters, even
within the realm of the standard model.


\subsection{Cosmic microwave background}
\label{sec:cmb}

The Cosmic Microwave Background Radiation (CMBR),
which is a direct relic of the early Universe, currently
provides the deepest probe of cosmological structures
and imposes severe constraints
on the various proposed matter evolution scenarios
and cosmological parameters.
Although the CMBR is a unique and deep probe of both
the thermal history of the early Universe and the
primordial perturbations in the matter distribution,
the associated anisotropies are not exclusively primordial
in nature.
Important modifications to the CMBR spectrum 
can arise from large scale coherent structures, even well after
the photons decouple from the matter at redshift $z\sim 10^3$,
due to the gravitational redshifting of the
photons through the Sachs-Wolfe effect arising from potential
gradients~\cite{SW67, AMTC91}
\begin{equation}
  \frac{\Delta T}{T} = \Phi_{\rm e} - \Phi_{\rm r}
  - \int_{\rm r}^{\rm e} \frac{2\vec{l}\cdot\nabla\Phi}{a} dt,
\end{equation}
where the integral is evaluated from the emission (e)
to reception (r) points along the spatial photon paths
$\vec{l}$, $\Phi$ is the gravitational potential,
$\Delta T/T$ defines the temperature fluctuations,
and $a(t)$ is the cosmological scale factor in the standard FLRW metric.
Also, if the intergalactic medium (IGM)
reionizes sometime after the decoupling,
say from an early generation of stars, the increased
rate of Thomson scattering off the
free electrons will erase sub-horizon scale
temperature anisotropies, while creating secondary
Doppler shift anisotropies.
To make meaningful comparisons between numerical models
and observed data, these effects 
(and others, see for example \S~\ref{sec:secondary}
and references~\cite{HSSW95, JL98}) must be 
incorporated self-consistently 
into the numerical models and to high accuracy
in order to resolve the weak signals.


\subsubsection{Ray-tracing}
\label{sec:raytracing}

Many computational analyses based on linear perturbation theory
have been carried out to estimate the temperature
anisotropies in the sky
(for example see~\cite{MB95} and the
references cited in~\cite{HSSW95}).
Although such linearized approaches yield reasonable
results, they are not well-suited to discussing the expected
imaging of the developing nonlinear structures in the
microwave background.
An alternative ray-tracing approach has been developed by
Anninos et al.~\cite{AMTC91} to introduce and propagate
individual photons through the evolving nonlinear matter
structures. They solve the geodesic equations of motion
and subject the photons to Thomson
scattering in a probabilistic way
and at a rate determined by the
local density of free electrons in the model.
Since the temperature fluctuations remain small,
the equations of motion for the photons are treated
as in the linearized limit, and the anisotropies are
computed according to
\begin{equation}
  \frac{\Delta T}{T} = \frac{\delta z}{1+z},
\end{equation}
where
\begin{equation}
  1+z = \frac{(k^\mu u_\mu)_{\rm e} }{ (k^\mu u_\mu)_{\rm r}},
\end{equation}
and the photon wave vector $k^\mu$ and
matter rest frame four-velocity $u_\mu$ are
evaluated at the emission (e) and reception (r) points.
Applying their procedure to a Hot Dark Matter (HDM) model
of structure formation,
Anninos et al.~\cite{AMTC91} find the parameters
for this model are severely constrained by COBE data 
such that $\Omega_0 h^2 \approx 1$,
where $\Omega_0$ and $h$ are the density and Hubble
parameters.


\subsubsection{Effects of reionization}
\label{sec:reionization}

In models where the IGM does not reionize,
the probability of scattering after the 
photon-matter decoupling epoch is low,
and the Sachs-Wolfe effect dominates the
anisotropies at angular scales larger than a few degrees.
However, if reionization occurs, the scattering probability
increases substantially and the matter structures, which 
develop large bulk motions relative to the comoving
background, induce Doppler shifts on the scattered
CMBR photons and leave an imprint of the surface of last scattering.
The induced fluctuations on subhorizon scales in
reionization scenarios can 
be a significant fraction of the primordial anisotropies,
as observed by Tuluie et al.~\cite{TMA95}.
They considered two possible scenarios of reionization:
A model that suffers
early and gradual (EG) reionization of the IGM
as caused by the photoionizing UV radiation emitted by
decaying neutrinos,
and the late and sudden (LS) scenario
as might be applicable to the case of an early generation
of star formation activity at high redshifts.
Considering the HDM model with $\Omega_0=1$ and $h=0.55$,
which produces CMBR anisotropies above current COBE limits
when no reionization is included (see \S~\ref{sec:raytracing}), 
they find that the
EG scenario effectively reduces the anisotropies
to the levels observed by COBE and generates smaller
Doppler shift anisotropies than the LS model, as demonstrated
in Figure~\ref{fig:cmb1}.
The LS scenario of reionization is not able to reduce
the anisotropy levels below the COBE limits,
and can even give rise to greater Doppler shifts
than expected at decoupling.

\begin{figure}[htbp]
  \def\epsfsize#1#2{1.1#1}
  \caption{\it The top two images represent temperature fluctuations
    (i.e., $\Delta T/T$) due to the Sachs-Wolfe effect  and Doppler
    shifts in a standard critically closed HDM model with no
    reionization  and baryon fractions 0.02 (plate 1:
    $4^\circ \times 4^\circ$, rms $ = 2.8\times10^{-5}$) and 0.2
    (plate 2: $8^\circ \times 8^\circ$, rms 
    $ = 3.4\times10^{-5}$). The bottom two plates image fluctuations in an
    ``early and gradual'' reionization scenario of decaying neutrinos
    with baryon fraction 0.02 (plate 3: $4^\circ \times 4^\circ$, rms
    $ = 1.3\times10^{-5}$; and plate 4: $8^\circ \times 8^\circ$, rms
    $ = 1.4\times10^{-5}$).}
  \label{fig:cmb1}
\end{figure}


\subsubsection{Secondary anisotropies}
\label{sec:secondary}

Additional sources of CMBR anisotropy can arise from the
interactions of photons with dynamically evolving matter
structures and nonstatic gravitational potentials.
Tuluie et al.~\cite{TLA96} considered the impact
of nonlinear matter condensations on the CMBR
in $\Omega_0 \le 1$ Cold Dark Matter (CDM) models,
focusing on the relative importance of secondary 
temperature anisotropies due to three different effects:
1) time-dependent variations
in the gravitational potential
of nonlinear structures as a result of collapse
or expansion (the Rees-Sciama effect); 
2) proper motion of nonlinear structures
such as clusters and superclusters across the sky;
and 3) the decaying gravitational potential 
effect from the evolution of perturbations in open models.
They applied the ray-tracing procedure of~\cite{AMTC91} to
explore the relative importance of these secondary 
anisotropies as a function of 
the density parameter $\Omega_0$ and
the scale of matter distributions.
They find that secondary temperature anisotropies are
dominated by the decaying potential effect at large scales,
but that all three sources of anisotropy can produce signatures
of order $\Delta T/T \sim 10^{-6}$ as shown in Figure~\ref{fig:cmb2}.

\begin{figure}[htbp]
  \def\epsfsize#1#2{1.1#1}
  \caption{\it The top two images represent the proper motion and
    Rees--Sciama effects in the CMBR for  a critically closed CDM model
    (upper left), together with the corresponding column density of
    voids and clusters over the same region (upper right).  The bottom
    two images show the secondary anisotropies dominated here by the
    decaying potential effect in an open cosmological model (bottom
    left), together with the corresponding gravitational potential
    over the same region (bottom right). The rms fluctuations in both
    cases are on the order of $\pm 5\times10^{-7}$, though the open
    model carries a somewhat larger signature.}
  \label{fig:cmb2}
\end{figure}

In addition to the effects discussed here,
many other sources of secondary
anisotropies (such as gravitational lensing,
the Vishniac effect accounting for matter velocities
and flows into local potential wells, and the Sunyaev-Zel'dovich
(\S~\ref{sec:szeffect})
distortions from the Compton scattering of CMB photons
by electrons in the hot cluster medium) can also be significant.
See reference~\cite{HSSW95} for a more 
complete list and thorough
discussion of the different sources of CMBR
anisotropies.


\subsection{Gravitational lensing}
\label{sec:lensing}

Observations of gravitational lenses~\cite{SEF92}
provide measures of the abundance and strength
of nonlinear potential fluctuations
along the lines of sight to distant objects.
Since these calculations are sensitive to the gravitational
potential, they may be more reliable than galaxy
and velocity field measurements as they are not
subject to the same ambiguities associated with biasing effects.
Also, since different cosmological models predict different
mass distributions, especially at the higher
redshifts, lensing calculations can potentially be used
to confirm or discard competing cosmological models.

As an alternative to solving
the computationally demanding lens equations,
Cen et al.~\cite{CGOT94}
developed an efficient scheme to
identify regions with surface densities capable of
generating multiple images accurately for
splittings larger than three arcseconds.
They applied this technique to a standard CDM model
with $\Omega_0=1$, and found that this model predicts
more large angle splittings ($> 8''$)
than are known to exist in the observed Universe.
Their results suggest that the standard CDM
model should be excluded as a viable model of our Universe.
A similar analysis for a flat low density CDM model
with a cosmological constant
($\Omega_0= 0.3$, $\Lambda/3H_0^2 = 0.7$) suggests a
lower and perhaps acceptable number of lensing events.
However, an uncertainty in their studies is the nature of
the lenses at and below the resolution of the numerical grid.
They model the lensing structures as simplified 
Singular Isothermal Spheres (SIS) with no
distinctive cores.

Large angle
splittings are generally attributed to larger structures
such as clusters of galaxies, and there are indications
that clusters have small but finite core radii
$r_{\rm core} \sim 20 - 30 \, h^{-1}$~kpc. Core radii of
this size can have an important effect on the 
probability of multiple imaging. 
Flores and Primack~\cite{FP95} 
considered the effects of assuming two 
different kinds of splitting sources:
isothermal spheres with small but finite
core radii $\rho \propto (r^2+r_{\rm core}^2)^{-1}$, 
and spheres with a Hernquist density
profile $\rho \propto r^{-1} (r+a)^{-3}$,
where $r_{\rm core} \sim 20 - 30 \, h^{-1}$~kpc and
$a \sim 300 \, h^{-1}$~kpc.
They find that the computed frequency
of large-angle splittings, when using the
nonsingular profiles, can potentially decrease by 
more than an order of 
magnitude relative to the SIS case and can bring
the standard CDM model into better agreement
with observations.
They stress that lensing events are sensitive
to both the cosmological model (essentially the number
density of lenses) and to the inner lens structure,
making it difficult to probe the models until the
structure of the lenses, both observationally
and numerically, is better known.


\subsection{First star formation}
\label{sec:psf}

In CDM cosmogonies, the fluctuation spectrum at
small wavelengths has a logarithmic dependence at mass scales smaller than
$10^8$ solar masses, which indicates that all small scale fluctuations in this
model collapse nearly simultaneously in time. This leads to very complex
dynamics during the formation of these first structures. Furthermore, the
cooling in these fluctuations is dominated by the rotational/vibrational modes
of hydrogen molecules that were able to form using the free electrons left
over from recombination and those produced by strong shock waves as
catalysts. The first structures to collapse may be capable of producing
pop~III stars and have a substantial influence on
the subsequent thermal evolution of the intergalactic medium,
as suggested by Figure~\ref{fig:reheating}, due to the
radiation emitted by the first generation stars
as well as supernova driven winds. To know the
subsequent fate of the Universe and which structures will survive or be
destroyed by the UV background, it is first necessary to know when
and how the first stars formed.

Ostriker and Gnedin~\cite{OG96} have carried out high resolution numerical
simulations of the reheating and reionization of the Universe due to
star formation bursts triggered by molecular hydrogen cooling.
Accounting for the chemistry of the primeval hydrogen/helium plasma,
self-shielding of the gas, radiative cooling, and a phenomenological
model of star formation, they find that two distinct star
populations form: the first generation pop~III from $H_2$ cooling
prior to reheating at redshift $z \ge 14$; and the second generation
pop~II at $z<10$ when the virial temperature of the gas clumps
reaches $10^4$ K and hydrogen line cooling becomes efficient.
Star formation slows in the intermittent epoch due to the 
depletion of $H_2$ by photo-destruction and reheating.
In addition, the objects which formed pop~III stars also initiate pop~II
sequences when their virial temperatures reach $10^4$ K
through continued mass accretion.

In resolving the details of a single star forming region in a CDM Universe,
Abel et al.~\cite{AANZ98, ABN00} 
implemented a non-equilibrium radiative cooling
and chemistry model~\cite{AAZN97, AZAN97} together with the hydrodynamics
and dark matter equations, evolving nine separate atomic and molecular species
(H, H$^+$, He, He$^+$, He$^{++}$, H$^-$, H$_2^+$, H$_2$, and e$^-$)
on nested and adaptively refined numerical grids.
They follow the collapse and fragmentation of primordial clouds
over many decades in mass and spatial dynamical range,
finding a core of mass $\sim 200~M_\odot$ forms from a halo of about
$\sim 10^5~M_\odot$ (where a significant number fraction of hydrogen
molecules are created) after less than one percent of the halo gas
cools by molecular line emission.
Bromm et al.~\cite{BCL99} use a different Smoothed Particle
Hydrodynamics (SPH) technique and a six species model
(H, H$^+$, H$^-$, H$_2^+$, H$_2$, and e$^-$)
to investigate the initial mass function
of the first generation pop~III stars. They evolve an isolated
$3 \sigma$ peak of mass $2\times10^6 M_\odot$ which collapses at
redshift $z\sim 30$ and forms clumps of mass $10^2 - 10^3 M_\odot$
which then grow by accretion and merging, suggesting
that the very first stars were massive and in agreement with~\cite{ABN00}.


\subsection{Lyman-alpha forest}
\label{sec:qals}

The Lyman-alpha forest represents the optically
thin (at the Lyman edge) component of
Quasar Absorption Systems (QAS), a collection of
absorption features in quasar spectra extending back to high
redshifts. QAS are effective probes of
the matter distribution and the
physical state of the Universe at early epochs when
structures such as galaxies are still forming and evolving.
Although stringent observational constraints have been placed on
competing cosmological models at large scales by
the COBE satellite and over the smaller scales of our local
Universe by observations of galaxies and clusters,
there remains sufficient flexibility
in the cosmological parameters that no single model has
been established conclusively. 
The relative lack of constraining observational
data at the intermediate to high redshifts 
($0 < z < 5$), where differences between
competing cosmological models are more pronounced,
suggests that QAS can potentially yield valuable
and discriminating observational data.

Several combined N-body and hydrodynamic numerical
simulations of the Lyman forest have been performed
recently (\cite{DHWK97, MCOR96, ZMAN97}, for example), and
all have been able to fit the observations remarkably well,
including the column density and Doppler width distributions, the
size of absorbers~\cite{CAZN97},
and the line number evolution.
Despite the fact that the cosmological models and parameters are
different in each case, the simulations give roughly similar
results provided that the proper
ionization bias is used ($b_{\rm ion}\equiv(\Omega_{\rm b} h^2)^2/\Gamma$,
where $\Omega_{\rm b}$ is the baryonic density
parameter, $h$ is the Hubble parameter and $\Gamma$ is the photoionization
rate at the hydrogen Lyman edge).
However, see~\cite{BMAN98} for a discussion of the sensitivity
of statistical properties on numerical resolution, and~\cite{MBMATNZ99} for a systematic comparison of 
five different cosmological models to determine which attributes
are sensitive physical probes or discriminators of models.
A theoretical paradigm has thus emerged
from these calculations in which Lyman-alpha absorption lines originate
from the relatively
small scale structure in pregalactic or intergalactic gas through
the bottom-up hierarchical formation picture in CDM-like Universes.
The absorption features originate in structures exhibiting a variety of
morphologies commonly found in numerical simulations 
(see Figure~\ref{fig:lyman}), 
including fluctuations in underdense regions, spheroidal
minihalos, and filaments extending over scales of a few megaparsecs.

\begin{figure}[htbp]
  \def\epsfsize#1#2{0.45#1}
  \caption{\it Distribution of the gas density at redshift $z=3$ from
    a numerical hydrodynamics simulation of the Lyman-alpha forest
    with a CDM spectrum normalized to second year COBE observations,
    Hubble parameter of $h=0.5$, a comoving box size of
    9.6\protect~Mpc, and baryonic density of $\Omega_{\rm b}=0.06$
    composed of 76\% hydrogen and 24\% helium.  The region shown is
    2.4\protect~Mpc (proper) on a side. The isosurfaces represent
    baryons at ten times the mean density and are color coded to the
    gas temperature (dark blue = $3\times 10^4$\protect~K, light blue
    = $3\times 10^5$\protect~K). The higher density contours trace out
    isolated spherical structures typically found at the intersections
    of the filaments. A single random slice through the cube is also
    shown, with the baryonic overdensity represented by a rainbow-like
    color map changing from black (minimum) to red (maximum). The
    He$^{+}$ mass fraction is shown with a wire mesh in this same
    slice. To emphasize fine structure in the minivoids, the mass
    fraction in the overdense regions has been rescaled by the gas
    overdensity wherever it exceeds unity.}
  \label{fig:lyman}
\end{figure}


\subsection{Galaxy clusters}
\label{sec:xray}

Clusters of galaxies are the largest gravitationally bound
systems known to be in quasi-equilibrium. This
allows for reliable estimates to be made of their mass
as well as their dynamical and thermal attributes. 
The richest clusters, arising from $3 \sigma$
density fluctuations, can be as
massive as $10^{14}$--$10^{15}$ solar masses, and the 
environment in these structures is composed of
shock heated gas with temperatures of order $10^7$--$10^8$ 
degrees Kelvin which emits
thermal bremsstrahlung and line radiation at X-ray energies.
Also, because of their spatial size of $\sim 1 \, h^{-1}$~Mpc
and separations of order $50 \, h^{-1}$~Mpc, they provide
a measure of nonlinearity on scales close to
the perturbation normalization scale $8 \, h^{-1}$~Mpc.
Observations of the substructure, distribution,
luminosity, and
evolution of galaxy clusters are therefore likely to provide
signatures of the underlying cosmology of our Universe,
and can be used as cosmological probes
in the observable redshift range $0\le z \le 1$.


\subsubsection{Internal structure}
\label{sec:internal}

Thomas et al.~\cite{TCCEFJNHPPW97}
investigated the internal structure of galaxy clusters
formed in high resolution N-body simulations
of four different cosmological models, including
standard, open, and flat but low density Universes.
They find that the structure of relaxed clusters is similar
in the critical and low density Universes, although
the critical density models contain
relatively more disordered clusters due to the
freeze-out of fluctuations in open Universes at late times.
The profiles of relaxed clusters are very similar
in the different simulations since most clusters are in
a quasi-equilibrium state inside the virial radius
and generally follow the universal density profile of 
Navarro et al.~\cite{NFW97}.
There does not appear to be a strong cosmological dependence
in the profiles as suggested by previous
studies of clusters formed from 
pure power law initial density fluctuations~\cite{CER94}.
However, because more young and dynamically evolving
clusters are found in critical density Universes, Thomas et al.
suggest that it may be possible to discriminate among 
the density parameters by looking for multiple cores in the
substructure of the dynamic cluster population. 
They note that a statistical
population of 20 clusters could distinguish
between open and critically closed Universes.


\subsubsection{Number density evolution}
\label{sec:number}

The evolution
of the number density of rich clusters of galaxies
can be used to compute $\Omega_0$ and $\sigma_8$ 
(the power spectrum normalization on scales of
$8 \, h^{-1}$~Mpc) when
numerical simulation results are combined with the
constraint $\sigma_8 \Omega_0^{0.5} \approx 0.5$,
derived from observed present-day abundances of rich clusters.
Bahcall et al.~\cite{BFC97}
computed the evolution of the cluster mass function
in five different cosmological model simulations and find that 
the number of high mass (Coma-like) clusters
in flat, low $\sigma_8$ models
(i.e., the standard CDM model with $\sigma_8\approx 0.5$)
decreases dramatically by a factor of approximately $10^3$
from $z=0$ to $z\approx 0.5$. For low $\Omega_0$,
high $\sigma_8$ models, the data result in a much
slower decrease in the number density of clusters
over the same redshift interval.
Comparing these results to
observations of rich clusters in the real Universe,
which indicate only a slight
evolution of cluster abundances to redshifts
$z\approx $0.5--1, they conclude that critically
closed standard CDM and Mixed Dark Matter (MDM) models
are not consistent with the observed data. The models
which best fit the data are the open models with
low bias ($\Omega_0 = 0.3\pm 0.1$ and $\sigma_8 = 0.85 \pm 0.5$),
and flat low density models with a cosmological constant
($\Omega_0=0.34\pm 0.13$ and $\Omega_0 + \Lambda = 1$).


\subsubsection{X-ray luminosity function}
\label{sec:luminosity}

The evolution of the X-ray luminosity function, as well as the number,
size and temperature distribution of galaxy clusters are all
potentially important discriminants of cosmological models and the
underlying initial density power spectrum that gives rise to these
structures. Because the X-ray luminosity (principally due to thermal
bremsstrahlung emission from electron/ion interactions in the hot
fully ionized cluster medium) is proportional to the square of the gas
density, the contrast between cluster and background intensities is
large enough to provide a window of observations that is especially
sensitive to cluster structure. Comparisons of simulated and observed
X-ray functions may be used to deduce the amplitude and shape of the
fluctuation spectrum, the mean density of the Universe, the mass
fraction of baryons, the structure formation model, and the background
cosmological model.

Several groups~\cite{BCNOS94, CO94} have examined the properties of
X-ray clusters in high resolution numerical simulations of a standard
CDM model normalized to COBE. Although the results are very sensitive
to grid resolution (see~\cite{AN96a} for a discussion of the effects
from resolution constraints on the properties of rich clusters), their
primary conclusion, that the standard CDM model predicts too many
bright X-ray emitting clusters and too much integrated X-ray
intensity, is robust since an increase in resolution will only
exaggerate these problems. On the other hand, similar calculations
with different cosmological models~\cite{CO94, BN97} suggest reasonable
agreement of observed data with Cold Dark Matter + cosmological
constant ($\Lambda$CDM), Cold + Hot Dark Matter (CHDM), and Open or
low density CDM (OCDM) evolutions due to different universal
expansions and density power spectra.


\subsubsection{SZ effect}
\label{sec:szeffect}

The Sunyaev-Zel'dovich (SZ) effect is the change in energy that
CMB photons undergo when they scatter in hot gas typically found
in cores of galaxy clusters. There are two main effects:
thermal and kinetic. Thermal SZ is the dominant mechanism
which arises from thermal motion of gas in the temperature
range $10^7$--$10^8$ K, and is described by the Compton $y$
parameter
\begin{equation}
  y = \sigma_{\rm T} \int \frac{n_{\rm e} k_{\rm B} T_{\rm e}}
  {m_{\rm e} c^2} dl,
\end{equation}
where $\sigma_{\rm T}=6.65\times10^{-25}$ cm$^2$ is the Thomson
cross-section, $m_{\rm e}$, $n_{\rm e}$ and $T_{\rm e}$ are the
electron rest mass, density and temperature, $c$ is the speed of
light, $k_{\rm B}$ is Boltzmann's constant, and the integration is
performed over the photon path.
Photon temperature anisotropies are related to the $y$ parameter by
$\Delta T/T \approx -2 y$ in the Rayleigh-Jeans limit.
The kinetic SZ effect is a less influential Doppler shift resulting
from the bulk motion of ionized gas relative to the rest frame
of the CMB.

Springel et al.~\cite{SWH00} used a Tree/SPH code to study
the SZ effects in a CDM cosmology with a cosmological constant. They
find a mean amplitude for thermal SZ ($ y = 3.8\times10^{-6}$)
just below current observed upper limits, and
a kinetic SZ about 30 times smaller in power.
Da~Silva et al.~\cite{SBLT99} compared thermal SZ maps in three
different cosmologies (low density + $\Lambda$, critical density,
and low density open model). Their results
are also below current limits: $y\approx 4\times10^{-6}$
for low density models with contributions from over a broad
redshift range $z\le 5$,
and $y \approx 1\times10^{-6}$ for the critical density model with
contributions mostly from $z<1$.
However, further simulations are needed to explore the
dependence of the SZ effect on microphysics, i.e., cooling,
star formation, supernovae feedback.


\subsection{Cosmological sheets}
\label{sec:sheets}

Cosmological sheets, or pancakes, form as overdense regions collapse
preferentially along one axis. Originally studied by
Zel'dovich~\cite{Zeldovich70} in the context of neutrino-dominated
cosmologies, sheets are ubiquitous features in nonlinear structure
formation simulations of CDM-like models with baryonic fluid, and
manifest on a spectrum of length scales and formation epochs. Gas
collapses gravitationally into flattened sheet structures, forming two
plane parallel shock fronts that propagate in opposite directions,
heating the infalling gas. The heated gas between the shocks then
cools radiatively and condenses into galactic structures. Sheets are
characterized by essentially five distinct components: the preshock
inflow, the postshock heated gas, the strongly cooling/recombination
front separating the hot gas from the cold, the cooled postshocked
gas, and the unshocked adiabatically compressed gas at the
center. Several numerical calculations~\cite{BCSW84, SS85, ANA95} have
been performed of these systems which include baryonic fluid with
hydrodynamical shock heating, ionization, recombination, dark matter,
thermal conductivity, and radiative cooling (Compton, bremsstrahlung,
and atomic line cooling), in both one and two spatial dimensions to
assert the significance of each physical process and to compute the
fragmentation scale. See also~\cite{AM99} where fully general
relativistic numerical calculations of cosmological sheets are
presented in plane symmetry, including relativistic hydrodynamical
shock heating and consistent coupling to spacetime curvature.

\begin{figure}[htbp]
  \def\epsfsize#1#2{0.35#1}
  \caption{\it Two different model simulations of cosmological sheets
    are presented: a six species model including only atomic line
    cooling (left), and a nine species model including also hydrogen
    molecules (right). The evolution sequences in the images show the baryonic
    overdensity and gas temperature at three redshifts following the
    initial collapse at $z=5$. In each figure, the vertical axis is
    32\protect~kpc long (parallel to the plane of collapse) and the
    horizontal axis  extends to 4\protect~Mpc on a logarithmic scale to
    emphasize the central structures. Differences in the two cases are
    observed in the cold pancake layer and the cooling flows between
    the shock front and the cold central layer. When the central layer
    fragments, the thickness of the cold gas layer in the six (nine)
    species case grows to 3 (0.3)\protect~kpc  and the surface density
    evolves with a dominant transverse mode corresponding to a scale
    of approximately 8 (1)\protect~kpc. Assuming a symmetric
    distribution of matter along the second transverse direction, the
    fragment masses are approximately $10^7$ ($10^5$) solar masses.}
  \label{fig:sheets}
\end{figure}

In addition, it is well known that gas which cools to 1~eV
through hydrogen line cooling will likely cool faster than it can
recombine. This nonequilibrium cooling increases the number of electrons and
ions (compared to the equilibrium case) which, in turn, increases the
concentrations of $H^-$ and $H_2^+$, the intermediaries that
produce hydrogen molecules $H_2$. If large
concentrations of molecules form, excitations of the vibrational/rotational
modes of the molecules can efficiently cool the gas to well below 1~eV, the
minimum temperature expected from atomic 
hydrogen line cooling. Because the gas
cools isobarically, the reduction in temperature results in an even greater
reduction in the Jeans mass, and the bound objects which form from the
fragmentation of $H_2$ cooled cosmological sheets may be associated with
massive stars or star clusters. 
Anninos and Norman~\cite{AN96b} have carried out 1D and 2D 
high resolution numerical
calculations to investigate the role of hydrogen molecules in the
cooling instability and fragmentation of cosmological sheets,
considering the collapse of perturbation wavelengths from
1~Mpc to 10~Mpc. They find that for the more energetic (long 
wavelength) cases, the mass fraction of hydrogen molecules
reaches $n_{\rm H2}/n_{\rm H} \sim 3\times10^{-3}$, which cools
the gas to $4\times10^{-3}$ eV and results in a fragmentation
scale of $9\times10^3~M_\odot$. This represents reductions
of 50 and $10^3$ in temperature and Jeans mass respectively when 
compared, as in Figure~\ref{fig:sheets}, to
the equivalent case in which hydrogen molecules were 
neglected.

However, the above calculations neglected
important interactions arising from self-consistent
treatments of radiation fields with ionizing
and photo-disso\-cia\-ting photons and self-shielding effects.
Susa and Umemura~\cite{SU00} studied the thermal history
and hydrodynamical collapse of pancakes in a UV background
radiation field. They solve the radiative transfer
of photons together with the hydrodynamics and chemistry
of atomic and molecular hydrogen species. Although their
simulations were restricted to one-dimensional plane
parallel symmetry, they suggest a classification scheme
distinguishing different dynamical behavior and
galaxy formation scenarios
based on the UV background radiation level and a 
critical mass corresponding to $1-2\sigma$ density
fluctuations in a standard CDM cosmology. These level
parameters distinguish galaxy formation scenarios
as they determine the local thermodynamics, the
rate of $H_2$ line emissions and cooling, the amount of
starburst activity, and the rate and mechanism of 
cloud collapse.

\newpage


\section{Conclusion}
\label{sec:conclusion}

This review is intended to provide
a flavor of the variety of numerical
cosmological calculations performed of different phenomena occurring
throughout the history of our Universe. The topics discussed
range from the strong field dynamical behavior of spacetime geometry
at early times near the Big Bang singularity and the epoch of inflation,
to the late time evolution of large scale
matter fluctuations and the formation of clusters of galaxies.
Although a complete, self-consistent, and accurate description
of our Universe is impractical considering the complex
multiscale and multiphysics requirements, a number of
enlightening results have been demonstrated through computations.
For example, both monotonic AVTD and chaotic oscillatory BLK
behavior have been found in the asymptotic approach to
the initial singularity in a small set of
inhomogeneous Bianchi and Gowdy models,
though it remains to be seen what the generic behavior
might be in more general multidimensional spacetimes.
Numerical calculations suggest that scalar fields
play an important complicated role in the nonlinear or chaotic
evolution of cosmological models with consequences
for the triggering (or not) of inflation and the subsequent
dynamics of structure formation. It is possible, for example, that
these fields can influence the details of inflation and
have observable ramifications as fractal patterns
in the density spectrum, gravitational waves, galaxy distribution,
and cosmic microwave background anisotropies.
All these effects require further studies. Numerical simulations
have been used to place limits on curvature anisotropies
and cosmological parameters
at early times by considering primordial nucleosynthesis
in anisotropic and inhomogeneous cosmologies.
Finally, the large collection of calculations performed of the
post-recombination epoch (for example, cosmic microwave, 
gravitational lensing, Lyman-alpha absorption, and galaxy cluster
simulations) have placed strong constraints
on the standard model parameters and structure formation
scenarios when compared to observations. Considering the range of
models consistent with inflation, the preponderance of 
observational, theoretical and computational data suggest a best fit
model that is spatially flat with a cosmological constant
and a small tilt in the power spectrum.

Obviously many fundamental issues remain unresolved, including
the background or topology of the cosmological model which best
describes our Universe, the generic singularity behavior,
the dynamics of inflaton fields, the imprint of complex
interacting scalar fields, the fundamental nonlinear curvature
and gravitational wave interactions,
the correct structure formation scenario,
and the origin and spectrum of primordial fluctuations, for example,
are uncertain. However the field of numerical cosmology has
matured in the development of computational techniques, the modeling
of microphysics, and in taking advantage of current computing
technologies, to the point that it is now possible to perform
high resolution multiphysics simulations and
reliable comparisons of numerical models with observed data.

\newpage


\section{Appendix: Basic Equations and Numerical Methods}
\label{sec:appendix}

Some basic equations relevant for 
fully relativistic as well as perturbative cosmological calculations
are summarized in this section,
including the complete Einstein equations, 
choices of kinematical conditions, initial data constraints,
stress-energy-momentum tensors, dynamical equations
for various matter sources, and the Newtonian counterparts
on background isotropic models. 
References to numerical methods are also provided.


\subsection{The Einstein equations}


\subsubsection{ADM formalism}
\label{sec:adm}

There are many ways to write the Einstein equations.
The most common is the ADM or 3\,+\,1
form~\cite{ADM62} which decomposes spacetime
into layers of three-dimensional space-like hypersurfaces,
threaded by a time-like normal congruence 
$n^\mu = (1, - \beta^i)/\alpha$.
The general spacetime metric is written as
\begin{equation}
  ds^2 = (-\alpha^2 + \beta_i \beta^i) dt^2 + 2\beta_i dx^i dt
  +\gamma_{ij} dx^i dx^j,
  \label{4metric}
\end{equation}
where $\alpha$ and $\beta^i$ are the lapse function and shift vector
respectively, and $\gamma_{ij}$ is the spatial 3-metric.
The lapse defines the proper time between consecutive layers of 
spatial hypersurfaces, the shift propagates the coordinate
system from 3-surface to 3-surface, and the induced 3-metric is
related to the 4-metric via 
$\gamma_{\mu\nu} = g_{\mu\nu} + n_\mu n_\nu$.
The Einstein equations are written as four constraint equations,
\begin{equation}
  ^{(3)}R + K^2 - K_{ij}K^{ij} = 16\pi G \rho_{\rm H},
  \label{eqn:ham}
\end{equation}
\vspace{-2 em}
\begin{equation}
  \nabla_i \left(K^{ij} - \gamma^{ij} K\right) = 8\pi G s^j,
  \label{eqn:mom} 
\end{equation}
twelve evolution equations,
{\setlength {\arraycolsep}{0.14 em}
\begin{equation}
  \begin{array}{rcl}
    \partial_t \gamma_{ij} &=& \displaystyle {\cal L}_\beta
    \gamma_{ij} - 2\alpha K_{ij}, \\
    \rule{0 cm}{0.4 cm}
    \partial_t K_{ij} &=& \displaystyle {\cal L}_\beta K_{ij} -
    \nabla_i \nabla_j \alpha + \\ 
    \rule{0 cm}{0.4 cm}
    & & \displaystyle \alpha\left[ ^{(3)}R_{ij} - 2K_{ik}K^k_j + K
    K_{ij} - 8\pi G \left(s_{ij} - \frac{1}{2} s \gamma_{ij} +
    \frac{1}{2} \rho_{\rm H} \gamma_{ij} \right)\right],
  \end{array}
  \label{eqn:dtkij}
\end{equation}}%
and four kinematical or coordinate conditions for the lapse
function and shift vector that can be specified arbitrarily
(see \S~\ref{sec:kinematic}). Here,
{\setlength {\arraycolsep}{0.14 em}
\begin{equation}
  \begin{array}{rcl}
    {\cal L}_\beta \gamma_{ij} &=& \nabla_i\beta_j +
    \nabla_j\beta_i, \\
    \rule{0 cm}{0.4 cm}
    {\cal L}_\beta K_{ij} &=& \beta^k \nabla_k K_{ij} + K_{ik}
    \nabla_j \beta^k + K_{kj} \nabla_i \beta^k,
  \end{array}
\end{equation}}%
where $\nabla_i$ is the spatial covariant derivative with
respect to $\gamma_{ij}$, $^{(3)}R_{ij}$ the
spatial Ricci tensor, $K$ the trace of the extrinsic
curvature $K_{ij}$, and $G$ is the gravitational constant.
The matter source terms
$\rho_{\rm H}$, $s^j$, $s_{ij}$ and $s=s^i_i$
as seen by the observers at rest in the time slices
are obtained from the appropriate projections
{\setlength {\arraycolsep}{0.14 em}
  \begin{eqnarray}
    \rho_{\rm H} &=& n^\mu n^\nu T_{\mu\nu}, \\
    s_i         &=& -\gamma^\mu_i n^\nu T_{\mu\nu}, \\
    s_{ij}      &=& \gamma^\mu_i \gamma^\nu_j T_{\mu\nu}
\end{eqnarray}}%
for the energy density, momentum density
and spatial stresses, respectively.
Here $c=1$, and
Greek (Latin) indices refer to 4(3)-dimensional quantities.

It is worth noting that several alternative formulations of 
Einstein's equations have been suggested, including hyperbolic
systems~\cite{Reula98} which have nice mathematical
properties, and conformal traceless systems~\cite{SN95, BS99} which
make use of a conformal
decomposition of the 3-metric and trace-free part of
the extrinsic curvature. Introducing
$\tilde{\gamma}_{ij} = e^{-4\psi}{\gamma}_{ij}$ with
$e^{4\psi} = \gamma^{1/3}$ so that the determinant of
$\tilde{\gamma}_{ij}$ is unity, and
$\tilde{A}_{ij} = e^{-4\psi}{A}_{ij}$, evolution equations
can be written in the conformal traceless system
for $\psi$, $\tilde{\gamma}_{ij}$, $K$,
$\tilde{A}_{ij}$ and the conformal connection functions,
though not all of these variables are independent.
However, it is not yet 
entirely clear which of these methods is best suited for generic
problems. For example, hyperbolic forms are easier to
characterize mathematically than ADM and may potentially
be more stable, but can suffer from greater inaccuracies
by introducing additional equations or higher order
derivatives. Conformal treatments are considered
to be generally more stable~\cite{BS99}, but can be less
accurate than traditional ADM for short term evolutions~\cite{ABDFPSST00}.

Many numerical methods have been used to solve the Einstein
equations, including 
variants of the leapfrog scheme, the method of McCormack, 
the two-step Lax-Wendroff method, and the iterative
Crank-Nicholson scheme, among others.
For a discussion and comparison of the different methods,
the reader is referred to~\cite{BHS89},
where a systematic study was carried out on spherically symmetric
black hole spacetimes using traditional ADM, and 
to~\cite{BS99, ABDFPSST00, AMSST97} (and references therein) which
discuss the stability and accuracy of hyperbolic and conformal
treatments.


\subsubsection{Symplectic formalism}
\label{sec:symplectic}

A different approach to conventional (i.e., 3\,+\,1 ADM) techniques
in numerical cosmology has been developed by 
Berger and Moncrief~\cite{BM93}. 
For example, they consider Gowdy cosmologies on $T^3\times R$
with the metric
{\setlength {\arraycolsep}{0.14 em}
\begin{eqnarray}
  ds^2 &=& e^{-\lambda/2} e^{\tau/2} \left(-e^{-2\tau} d\tau^2
    + d\theta^2\right) + \nonumber \\
  & & e^{-\tau} \left[e^P d\sigma^2
    + 2e^P Q d\sigma d\delta + \left(e^P Q^2 + e^{-P}\right)
    d\delta^2 \right],
  \label{eqn:gowdy_met}
\end{eqnarray}}%
where $\lambda$, $P$ and $Q$ are functions of $\theta$ and $\tau$,
and the coordinates are bounded by
$0\le (\theta,\sigma,\delta) \le 2\pi$. The singularity 
corresponds to the limit $\tau \rightarrow \infty$.
For small amplitudes, $P$ and $Q$ may be identified with $+$ and
$\times$ polarized gravitational wave components and
$\lambda$ with the background cosmology through which
they propagate.
An advantage of this formalism is that the initial value
problem becomes trivial since $P$, $Q$ and their first
derivatives may be specified arbitrarily
(although it is not quite so trivial in more general spacetimes).

Although the resulting Einstein equations
can be solved in the usual spacetime discretization fashion,
an interesting alternative
method of solution is the symplectic operator splitting
formulation~\cite{BM93, Moncrief83}
founded on recognizing that the
second order equations can be obtained from the variation of a
Hamiltonian decomposed into kinetic and potential subhamiltonians,
\begin{equation}
  H = H_1 + H_2 =
  \frac12\oint_0^{2\pi} d\theta \left(\pi^2_P +
  e^{-2P}\pi_Q^2\right)+\frac12\oint_0^{2\pi} d\theta e^{-2\tau}
  \left(P^2_{,\theta} + e^{2P}Q_{,\theta}^2\right).
\end{equation}
The symplectic method 
approximates the evolution operator by
\begin{equation}
  e^{H\Delta \tau} = 
  e^{H_2\Delta \tau/2} e^{H_1\Delta \tau} e^{H_2\Delta \tau/2}
  + {\cal O}(\Delta\tau)^3,
\end{equation}
although higher order representations are possible.
If the two Hamiltonian components $H_1$ and $H_2$ are each
integrable, their solutions can be substituted directly into
the numerical evolution to provide potentially more accurate
solutions with fewer time steps~\cite{Berger95}.
This approach is well-suited for studies of singularities if
the asymptotic behavior is determined primarily by the kinetic
subhamiltonian, a behavior referred to as 
Asymptotically Velocity Term Dominated (see \S~\ref{sec:avtd}
and~\cite{Berger98}).

Symplectic integration methods are applicable to other spacetimes.
For example, Berger et al.~\cite{BGS96} developed a variation of this
approach to explicitly take advantage
of exact solutions for scattering between Kasner epochs
in Mixmaster models. Their algorithm evolves Mixmaster spacetimes
more accurately with larger time steps than previous methods.


\subsection{Kinematic conditions}
\label{sec:kinematic}

For cosmological simulations,
one typically takes the shift vector to be zero, hence
${\cal L}_\beta \gamma_{ij} = {\cal L}_\beta K_{ij} = 0$.
However, the shift can be used advantageously in deriving conditions
to maintain the 3-metric in a particular form, and
to simplify the resulting differential equations~\cite{CW83, CW84}.
See also reference~\cite{Shibata99} describing an approximate
minimum distortion gauge condition used to help stabilize
simulations of general relativistic 
binary clusters and neutron stars.

Several options can be implemented for the lapse function,
including geodesic ($\alpha=1$), algebraic, and
mean curvature slicing.
The algebraic condition takes the form
\begin{equation}
  \alpha = F_1(x^\mu) F_2(\gamma),
  \label{eqn:algebraic}
\end{equation}
where $F_1(x^\mu)$ is an arbitrary function of coordinates $x^\mu$,
and $F_2(\gamma)$ is a dynamic function of the determinant of
the 3-metric. This choice is computationally cheap,
simple to implement, and
certain choices of $F_2$ (i.e., $1+\ln\gamma$) can 
mimic maximal slicing in its singularity avoidance
properties~\cite{ACMSST95}.
On the other hand, numerical solutions derived from harmonically-sliced
foliations can exhibit pathological gauge behavior in the
form of coordinate ``shocks'' or singularities which will
affect the accuracy, convergence and stability
of solutions~\cite{Alcubierre97, Hern00}. Also,
evolutions in which the lapse function is fixed by some
analytically prescribed method (either geodesic or near-geodesic) can
be unstable, especially for sub-horizon scale
perturbations~\cite{Anninos98}.

The mean curvature slicing equation is derived by taking the trace 
of the extrinsic curvature evolution equation~(\ref{eqn:dtkij}),
\begin{equation}
  \nabla^i \nabla_i \alpha = \alpha\left[K_{ij} K^{ij} 
  + 4\pi G \left(\rho_{\rm H} + s\right)\right] 
  +\beta^i \nabla_i K - \partial_t K,
  \label{eqn:maximal}
\end{equation}
and assuming $K=K(t)$, which can either be specified arbitrarily
or determined by imposing a boundary condition on the lapse
function after solving~(\ref{eqn:maximal}) for the quantity
$\alpha/\partial_t K$~\cite{CW84}.
It is also useful to consider replacing $\partial_t K$
in equation~(\ref{eqn:maximal}) with an exponentially driven
form as suggested by Eppley~\cite{Eppley79}, to reduce
gauge drifting and numerical errors in maximal~\cite{BDSSTW96} and mean curvature~\cite{Anninos98} sliced spacetimes.
The mean curvature slicing condition is the most natural one for 
cosmology as it foliates homogeneous cosmological spacetimes with
surfaces of homogeneity. 
Also, since
$K$ represents the convergence of coordinate curves from
one slice to the next and if it is constant, then localized
caustics (crossing of coordinate curves) 
and true curvature singularities can be avoided.
Whether general inhomogeneous 
spacetimes can be foliated with constant
mean curvature surfaces remains unknown. However,
for Gowdy spacetimes with two Killing fields and topology
$T^3\times R$, Isenberg and Moncrief~\cite{IM82} 
proved that such foliations
do exist and cover the entire spacetime.


\subsection{Sources of matter}
\label{sec:matter}


\subsubsection{Cosmological constant}
\label{sec:cosmocon}

A cosmological constant is implemented 
in the 3\,+\,1 framework simply by
introducing the quantity $-\Lambda/(8\pi G)$ 
as an effective isotropic pressure in the stress-energy tensor
\begin{equation}
  T_{\mu\nu} = -\frac{\Lambda}{8\pi G} ~g_{\mu\nu} .
\end{equation}
The matter  source terms
can then be written
{\setlength {\arraycolsep}{0.14 em}
\begin{eqnarray}
  \rho_{\rm H} &=& \frac{\Lambda}{8\pi G}, \\
  s_{ij} &=& - \frac{\Lambda}{8\pi G}\gamma_{ij},
\end{eqnarray}}%
with $s^i=0$.


\subsubsection{Scalar field}
\label{sec:scalar}

The dynamics of scalar fields is governed by the Lagrangian density
\begin{equation}
  {\cal{L}} = - \frac{1}{2} \sqrt{-g} \left[ g^{\mu\nu} \phi_{;\mu}
  \phi_{;\nu} + \xi R f(\phi) + 2 V(\phi) \right],
\end{equation}
where $R$ is the scalar Riemann curvature, $V(\phi)$ is the interaction
potential, $f(\phi)$ is typically assumed to be $f(\phi) = \phi^2$,
and $\xi$ is the field-curvature coupling constant
($\xi=0$ for minimally coupled fields and $\xi=1/6$ for
conformally coupled fields).
Varying the action yields the Klein-Gordon equation
\begin{equation}
  g^{\mu\nu} \phi_{;\mu\nu} - \xi R\phi - \partial_\phi V(\phi) = 0,
  \label{eqn:kleingordon}
\end{equation}
for the scalar field and
{\setlength {\arraycolsep}{0.14 em}
\begin{eqnarray}
  T_{\mu\nu} &=& (1-2\xi) \phi_{;\mu} \phi_{;\nu} 
  + \left(2\xi - \frac12\right) g_{\mu\nu}
  \phi_{;\sigma}\phi^{;\sigma} \nonumber \\
  & & - 2\xi\phi\phi_{;\mu\nu}
  + 2\xi\phi g_{\mu\nu} g^{\sigma\lambda} \phi_{;\sigma\lambda}
  + \xi G_{\mu\nu} \phi^2 - g_{\mu\nu} V(\phi),
\end{eqnarray}}%
for the stress-energy tensor, where $G_{\mu\nu}=R_{\mu\nu}-g_{\mu\nu}R/2$.

For a massive, minimally coupled scalar field~\cite{BD82}
\begin{equation}
  T_{\mu\nu} = \phi_{;\mu} \phi_{;\nu} 
  - \frac12 g_{\mu\nu} g^{\rho\sigma}\phi_{;\rho}\phi_{;\sigma}
  - g_{\mu\nu} V(\phi),
\end{equation}
and
{\setlength {\arraycolsep}{0.14 em}
\begin{eqnarray}
  \rho_{\rm H} &=& \frac{1}{2}\gamma^{ij} \phi_{;i} \phi_{;j}
  +\frac{1}{2}\eta^2 +V(\phi), \\
  s_i    &=&-\eta \phi_{;i}, \\
  s_{ij} &=&\gamma_{ij}\left(-\frac{1}{2} \gamma^{kl} \phi_{;k}
  \phi_{;l}+\frac{1}{2}\eta^2 - V(\phi)\right) +\phi_{;i} \phi_{;j},
\end{eqnarray}}%
where
\begin{equation}
  \eta = n^\mu\partial_\mu \phi = \frac{1}{\alpha} (\partial_t -
  \beta^k\partial_k)\phi,
  \label{eqn:scalar_eta}
\end{equation}
$n^\mu = (1, - \beta^i)/\alpha$,
and $V(\phi)$ is a general potential which, for example,
can be set to $V= \lambda\phi^4$ in the chaotic inflation model.
The covariant form of the scalar field equation~(\ref{eqn:kleingordon})
can be expanded as in~\cite{KLM93} to yield
\begin{equation}
  \frac{1}{\alpha} \left(\partial_t - \beta^k \partial_k\right)\eta =
  \frac{1}{\sqrt{\gamma}} \partial_i (\sqrt{\gamma} \gamma^{ij}
  \partial_j\phi) + \frac{1}{\alpha} \gamma^{ij} \partial_i \alpha
  \partial_j\phi + K\eta - \partial_\phi V(\phi),
\end{equation}
which, when coupled to~(\ref{eqn:scalar_eta}), determines the
evolution of the scalar field.


\subsubsection{Collisionless dust}
\label{sec:dust}

The stress-energy tensor for a fluid composed of
collisionless particles (or dark matter)
can be written simply as the sum of the
stress-energy tensors for each particle~\cite{Weinberg72},
\begin{equation}
  T_{\mu\nu} = \sum m n u_\mu u_\nu,
  \label{tmunu}
\end{equation}
where $m$ is the rest mass of the particles, $n$ is the number density
in the comoving frame, and $u^\mu$ is the 4-velocity of each particle.
The matter source terms are
{\setlength {\arraycolsep}{0.14 em}
\begin{eqnarray}
  \rho_{\rm H} &=& \sum m n (\alpha u^0)^2, \\
  s_i         &=& \sum mn u_i (\alpha u^0), \\
  s_{ij}      &=& \sum mn u_i u_j~.
\end{eqnarray}}%
There are two conservation laws: the conservation of particles
$\nabla_\mu (n u^\mu) =0$,
and the conservation of energy-momentum
$\nabla_\mu T^{\mu \nu}=0$,
where $\nabla_\mu$ is the covariant derivative in the full 4-dimensional
spacetime. 
Together these conservation laws lead to
$u^\mu \nabla_\mu u^\nu =0$,
the geodesic equations of motion for the particles, which can be 
written out more explicitly in the
computationally convenient form
{\setlength {\arraycolsep}{0.14 em}
\begin{eqnarray}
  \frac{dx^i}{dt} &=& \frac{g^{i\alpha} u_\alpha}{u^0}, \\
  \frac{du_i}{dt} &=& -\frac{u_\alpha u_\beta \partial_i
  g^{\alpha \beta}}{2 u^0},
\end{eqnarray}}%
where $x^i$ is the coordinate position of each particle,
$u^0$ is determined by the normalization $u^\mu u_\mu = -1$,
\begin{equation}
  \frac{d}{dt} \equiv v^\mu \partial_\mu = \partial_t + v^i \partial_i
\end{equation}
is the Lagrangian derivative, and
$v^\mu = u^\mu / u^0$ is the ``transport'' velocity
of the particles as measured by
observers at rest with respect to the coordinate grid.


\subsubsection{Ideal gas}
\label{sec:hydro}

The stress-energy tensor for a perfect fluid is
\begin{equation}
  T_{\mu\nu} = \rho h u_\mu u_\nu + P g_{\mu\nu},
  \label{eq:Th}
\end{equation}
where $g_{\mu\nu}$ is the 4-metric,
$h = 1 + \epsilon + {P / \rho}$ is the relativistic enthalpy,
and $\epsilon$, $P$, $\rho$ and $u_\mu$ are the specific internal
energy (per unit mass), pressure, rest mass density and four-velocity
of the fluid. Defining
\begin{equation}
  u = -n_\mu u^\mu = \alpha u^0 = \left(1+u^i u_i\right)^{1/2}
    = \left(1-\frac{V_i V^i}{\alpha^2}\right)^{-1/2},
  \label{eqn:boost}
\end{equation}
as the generalization of the special relativistic boost factor,
the matter source terms become
{\setlength {\arraycolsep}{0.14 em}
\begin{eqnarray}
  \rho_{\rm H} &=& \rho h u^2 - P,  \\
  s_i         &=& \rho h u u_i, \\
  s_{ij}      &=& P \gamma_{ij} + \rho h u_i u_j .
\end{eqnarray}}%
The hydrodynamics equations are derived from the
normalization of the 4-velocity, $u^{\mu}u_{\mu} = -1$,
the conservation of baryon number, $\nabla_{\mu} (\rho u^{\mu}) = 0$,
and the conservation of energy-momentum, $\nabla_{\mu} T^{\mu\nu} = 0$.
The resulting equations can be written in flux conservative 
form as~\cite{Wilson79}
\begin{equation}
  {\partial D \over \partial t} +
  {\partial (D V^i) \over \partial x^i} = 0,
  \label{eqn:denseqn}
\end{equation}
\begin{equation}
  {\partial E \over \partial t} + {\partial (E V^i) \over \partial
  x^i} + P {\partial W \over \partial t} + P {\partial (W V^i) \over
  \partial x^i} = 0,
  \label{eqn:eneqn}
\end{equation}
\begin{equation}
  \frac{\partial S_i}{\partial t} +
  \frac{\partial (S_i V^j)}{\partial x^j} -
  \frac{S^\mu S^\nu}{2S^0} \frac{\partial g_{\mu\nu}}{\partial x^i} +
  \sqrt {-g} \frac{\partial P}{\partial x^i} = 0,
  \label{eqn:momeqn}
\end{equation}
where $W=\sqrt{-g} u^0$, $D=W \rho$, $E=W \rho \epsilon$,
$S_i = W \rho h u_i$, $V^i=u^i/u^0$,
and $g$ is the determinant of the 4-metric satisfying
$\sqrt{-g}  = \alpha\sqrt{\gamma}$.
A prescription for specifying an equation of state
(e.g., $P=(\Gamma-1) E/W = (\Gamma-1)\rho\epsilon$ for an ideal gas)
completes the above equations.

When solving 
equations~(\ref{eqn:denseqn}, \ref{eqn:eneqn}, \ref{eqn:momeqn}),
an artificial viscosity method is needed 
to handle the formation and propagation of shock
fronts~\cite{Wilson79, HSW84a, HSW84b}. Although these
methods are simple to implement and are computationally
efficient, they are inaccurate for ultrarelativistic
flows with very high Lorentz factors.
On the other hand, a number of different
formulations~\cite{Font00}
of these equations have been developed to
take advantage of the hyperbolic and
conservative nature of the equations in using high resolution
shock capturing schemes (although strict conservation is only possible
in flat spacetimes -- curved spacetimes exhibit source terms due to
geometry). 
Such high resolution Godunov techniques~\cite{RIMM96, BFIMM97}
provide more accurate and stable treatments
in the highly relativistic regime.
A particular formulation due to~\cite{BFIMM97} is the following:
\begin{equation}
  \frac{\partial \sqrt{\gamma}~U  (\vec{w})}{\partial t} +
  \frac{\partial \sqrt{-g    }~F^i(\vec{w})}{\partial x^i} 
  = \sqrt{-g}~S(\vec{w}),
  \label{eqn:fluxcons}
\end{equation}
where
{\setlength {\arraycolsep}{0.14 em}
\begin{eqnarray}
  S(\vec{w}) &=& \!\left[0, T^{\mu\nu} \left(\frac{\partial
  g_{\nu j}}{\partial x^\mu} - \Gamma^\delta_{\nu\mu}
  g_{\delta j}\right)\!\!, \alpha\left(T^{\mu 0} \frac{\partial
  \ln\alpha}{\partial x^\mu} - T^{\mu\nu} \Gamma^0_{\nu\mu} \right)
  \right]\!,  \\
  F^i(\vec{w}) &=& \!\left[D\!\left(v^i\!-\!\frac{\beta^i}
  {\alpha}\right)\!\!, S_j\!\left(v^i\!-\!\frac{\beta^i} 
  {\alpha}\right)\! + P\delta^i_j,
  (E\!-\!D)\!\left(v^i\!-\!\frac{\beta^i}
  {\alpha}\right)\! + Pv^i\right]\!,
\end{eqnarray}}%
and
$\vec{w}    = (\rho, v_i, \epsilon)$,
$U(\vec{w}) = (D, S_i, E-D)$,
$E          =\rho h \tilde{W}^2 - P$,
$S_j        = \rho h \tilde{W}^2 v_j$,
$D          =\rho \tilde{W}$,
$v^i        = \gamma^{ij} v_j = u^i/(\alpha u^0) + \beta^i/\alpha$,
and 
$\tilde{W}  = \alpha u^0 = (1 - \gamma_{ij} v^i v^j)^{-1/2}$.


\subsection{Constrained nonlinear initial data}
\label{sec:initial}

One cannot take arbitrary data to initiate an evolution
of the Einstein equations.
The data must satisfy the constraint equations~(\ref{eqn:ham})
and~(\ref{eqn:mom}). York~\cite{York79} developed
a procedure to generate proper initial data by introducing 
conformal transformations of the 3-metric
$\gamma_{ij} = \psi^4\hat\gamma_{ij}$, the
trace-free momentum components 
$A^{ij}=K^{ij}-\gamma^{ij}K/3 = \psi^{-10}\hat A^{ij}$,
and matter source terms $s^i=\psi^{-10}\hat s^i$ and
$\rho_{\rm H}= \psi^{-n}\hat\rho_{\rm H}$, where $n>5$ for uniqueness
of solutions to the elliptic equation~(\ref{eqn:ham2}) below.
In this procedure, the conformal (or ``hatted'')
variables are freely specifiable.
Further decomposing the free momentum variables into transverse
and longitudinal components
$\hat A^{ij} = \hat A^{ij}_* + (\hat l w)^{ij}$,
the Hamiltonian and momentum constraints are written as
\begin{equation}
  \hat\nabla_i \hat\nabla^i \psi -\frac{\hat R}{8}\psi + \frac{1}{8}
  \hat A_{ij} \hat A^{ij}\psi^{-7} - \frac{1}{12} K^2 \psi^5 + 2\pi G
  \hat\rho \psi^{5-n} = 0,
  \label{eqn:ham2}
\end{equation}
\vspace{-1 em}
\begin{equation}
  (\hat \nabla_j \hat \nabla^j w)^i + \frac{1}{3} \hat \nabla^i
  \left(\hat \nabla_j w^j\right) + \hat R^i_j w^j - \frac{2}{3} \psi^6
  \hat \nabla^i K - 8\pi G \hat s^i = 0,
  \label{eqn:mom2}
\end{equation}
where the longitudinal part of $\hat A^{ij}$ is
reconstructed from the solutions by
\begin{equation}
  (\hat l w)^{ij} = \hat \nabla^i w^j + \hat \nabla^j w^i 
  -\frac{2}{3} \hat \gamma^{ij} \hat \nabla_k w^k.
\end{equation}
The transverse part of $\hat A^{ij}$ is constrained to satisfy 
$\hat\nabla_j \hat A^{ij}_* = \hat A^{\ j}_{*j} = 0$.

Equations~(\ref{eqn:ham2}) and~(\ref{eqn:mom2}) form a
coupled nonlinear set of elliptic equations which must
be solved iteratively, in general.
The two equations can, however,
be decoupled if a mean curvature slicing ($K=K(t)$) is assumed.
Given the free data $K$, $\hat\gamma_{ij}$, $\hat s^i$
and $\hat\rho$, the constraints are solved for
$\hat A^{ij}_*$, $(\hat l w)^{ij}$ and $\psi$.
The actual metric $\gamma_{ij}$ 
and curvature $K_{ij}$ are then reconstructed 
by the corresponding conformal transformations to provide
the complete initial data.
Reference~\cite{Anninos98} describes a procedure 
using York's formalism to
construct parametrized inhomogeneous initial data in
freely specifiable background spacetimes with matter sources.
The procedure is general enough to
allow gravitational wave and Coulomb nonlinearities
in the metric, longitudinal momentum fluctuations, 
isotropic and anisotropic background spacetimes,
and can accommodate the conformal-Newtonian
gauge to set up gauge invariant cosmological perturbation
solutions as free data.


\subsection{Newtonian limit}
\label{sec:newtonian}

The Newtonian limit is defined by spatial scales much smaller
than the horizon radius, peculiar velocities small compared to
the speed of light, and a gravitational potential that is both 
much smaller than unity (in geometric units)
and slowly varying in time.
A comprehensive review of the theory of cosmological
perturbations can be found in~\cite{MFB92}.


\subsubsection{Dark and baryonic matter equations}
\label{sec:matterequations}

The appropriate perturbation equations in this limit
are easily derived for a background FLRW expanding model,
assuming a metric of the form
\begin{equation}
  ds^2 = -(1+2\Phi) dt^2 + a(t)^2(1-2\Phi)\gamma_{ij} dx^i dx^j,
  \label{eqn:metric_flrw}
\end{equation}
where
\begin{equation}
  \gamma_{ij} = \delta^i_j \left(1 + \frac{k r^2}{4}
  \right)^{-2}\!\!\!\!\!\!\!,
\end{equation}
and $k = -1,~0,~+1$ for open, flat and closed Universes.
Also, $a \equiv 1/(1+z)$ is the 
cosmological scale factor, $z$ is the redshift, and $\Phi$ is the
comoving inhomogeneous gravitational potential.

The governing equations in the Newtonian limit are the 
hydrodynamic conservation equations,
\begin{equation}
  \frac{\partial {\tilde{\rho}_{\rm b}}}{\partial t} 
  + {\frac{\partial}{\partial x^i}}({\tilde{\rho}_{\rm b}}{v_{\rm b}^i})
  + 3{\frac{\dot a}{a}}{\tilde{\rho}_{\rm b}} = 0,
  \label{hydromass}
\end{equation}
\begin{equation}
  \frac{\partial ({\tilde{\rho}_{\rm b}}{{v}_{\rm b}^j})} {\partial t}
  + {\frac{\partial}{\partial x^i}} ({\tilde{\rho}_{\rm b}}
  {{v}_{\rm b}^i}~{v}_{\rm b}^j) + 5{\frac{\dot{a}}{a}}
  {\tilde{\rho}_{\rm b}}{{v}_{\rm b}^j} + {\frac{1}{a^2}}
  {\frac{\partial \tilde{p}}{\partial x^j}} +
  {\frac{\tilde{\rho}_{\rm b}}{a^2}}{\frac{\partial\tilde{\Phi}}
  {\partial x^j}} = 0,
  \label{hydromom}
\end{equation}
\begin{equation}
  \frac{\partial \tilde{e}}{\partial t} 
  + {\frac{\partial}{\partial x^i}} (\tilde{e} {{v}_{\rm b}^i})
  + \tilde{p}~\frac{\partial v^i_{\rm b}}{\partial x^i}
  + 3\frac{\dot a}{a}(\tilde{e} + \tilde{p})
  = \tilde{S}_{\rm cool},
  \label{hydroenergy}
\end{equation}
the geodesic equations for collisionless dust or dark matter 
(in comoving coordinates),
\begin{equation}
  \frac{d{x_{\rm d}^i}}{dt} = {{v}_{\rm d}^i}, 
\end{equation}
\begin{equation}
  \frac{d{{v}_{\rm d}^i}}{dt} = - 2{\frac{\dot{a}}{a}} {{v}_{\rm d}^i}
  - \frac{1}{a^2} \frac{\partial\tilde{\Phi}}{\partial x^i},
\end{equation}
Poisson's equation for the gravitational potential,
\begin{equation}
  {\nabla^2}{\tilde{\Phi}} = 4{\pi}G{a^2}(\tilde{\rho}_{\rm b} +
  \tilde{\rho}_{\rm d} - \tilde{\rho}_0),
\end{equation}
and the Friedman equation for the cosmological scale factor,
\begin{equation}
  \frac{d a}{dt} = H_0 \left[ \Omega_{\rm m} (\frac{1}{a} -1) +
  \Omega_\Lambda (a^2 -1) + 1 \right]^{1/2}.
\end{equation}
Here $\tilde{\rho}_{\rm d}$, $\tilde{\rho}_{\rm b}$, $\tilde{p}$ and
$\tilde{e}$ are the dark matter density, baryonic density, pressure
and internal energy density in the proper reference frame, $x^i$ and
${v}_{\rm b}^i$ are the baryonic comoving coordinates and peculiar
velocities, $x_{\rm d}^i$ and ${v}_{\rm d}^i$ are the dark matter
comoving coordinates and peculiar velocities,
$\tilde{\rho}_0=3H_0^2\Omega_0/(8\pi G a^3)$ is the proper background
density of the Universe, $\Omega_0$ is the total density parameter,
$\Omega_{\rm m}=\Omega_{\rm b} + \Omega_{\rm d}$ is the density
parameter including both baryonic and dark matter contributions,
$\Omega_\Lambda= \Lambda /(3H_0^2)$ is the density parameter
attributed to the cosmological constant $\Lambda$,
$H_0=100 \, h {\rm\ km\ s}^{-1}{\rm\ Mpc}^{-1}$ is the present Hubble
constant with $0.5<h<1$, and $\tilde{S}_{\rm cool}$ represents
microphysical radiative cooling and heating rates which can include
Compton cooling (or heating) due to interactions of free electrons
with the CMBR, bremsstrahlung, and atomic and molecular line
cooling. Notice that `tilded' (`non-tilded') variables refer to
proper (comoving) reference frame attributes.

An alternative total energy conservative form
of the hydrodynamics equations that allows high
resolution Godunov-type shock capturing techniques is
\begin{equation}
  \frac{\partial {{\rho}_{\rm b}}}{\partial t} +
  \frac{1}{a}{\frac{\partial}{\partial x^i}}
  ({{\rho}_{\rm b}}{\tilde{v}_{\rm b}^i}) = 0,
  \label{hydromass2}
\end{equation}
\begin{equation}
  \frac{\partial ({{\rho}_{\rm b}}{\tilde{v}_{\rm b}^j})}{\partial t}
  + \frac{1}{a}{\frac{\partial}{\partial x^i}} 
  ({{\rho}_{\rm b}}{\tilde{v}_{\rm b}^i}~\tilde{v}_{\rm b}^j +
  {p}\delta_{i j}) + \frac{\dot a}{a} {\rho}_{\rm b}
  \tilde{v}_{\rm b}^j + \frac{{\rho}_{\rm b}}{a} \frac{\partial
  \tilde{\Phi}}{\partial x^j} = 0,
  \label{hydromom2}
\end{equation}
\begin{equation}
  \frac{\partial {E}}{\partial t} 
  + \frac{1}{a}{\frac{\partial}{\partial x^i}} 
  ({E} \tilde{v}_{\rm b}^i + {p} \tilde{v}_{\rm b}^i)
  + \frac{2\dot a}{a} {E} + \frac{\rho_{\rm b} \tilde{v}_{\rm b}^i}{a}
  \frac{\partial \tilde{\Phi}}{\partial x^i} = {S}_{\rm cool},
  \label{hydroenergy2}
\end{equation}
with the corresponding particle and gravity equations
\begin{equation}
  \frac{d{x_{\rm d}^i}}{dt} = \frac{\tilde{v}_{\rm d}^i}{a}, 
\end{equation}
\begin{equation}
  \frac{d{\tilde{v}_{\rm d}^i}}{dt} = -
  {\frac{\dot{a}}{a}}{\tilde{v}_{\rm d}^i} - \frac{1}{a}
  \frac{\partial\tilde{\Phi}}{\partial x^i},
\end{equation}
\begin{equation}
  {{\nabla}^2}{\tilde{\Phi}} = \frac{4 \pi G}{a}(\rho_{\rm b} +
  \rho_{\rm d} - \rho_0),
\end{equation}
where ${\rho}_{\rm b}$ is the comoving density, 
${\rho}_0 = a^3 \tilde{\rho}_0$,
$\tilde{v}_{\rm b}^i$ is the proper frame peculiar velocity,
$p$ is the comoving
pressure, $E=\rho_b \tilde{v}_b^2/2 + p/(\gamma - 1)$ is the 
total peculiar energy per comoving volume, and $\tilde{\Phi}$ is the
gravitational potential.

These equations are easily extended~\cite{AZAN97} to include
reactive chemistry of nine separate atomic and molecular species
(H, H$^+$, He, He$^+$, He$^{++}$, H$^-$, H$_2^+$, H$_2$, and e$^-$),
assuming a common flow field, supplementing the 
total mass conservation equation~(\ref{hydromass}) with
\begin{equation}
  \frac{\partial \tilde{\rho}_j}{\partial t} 
  + \frac{\partial}{\partial x^i} (\tilde{\rho}_j {v}^i_{\rm b})
  + 3{\frac{\dot a}{a}}{\tilde{\rho}_j} 
  = {\sum_{i}}{\sum_{l}}{{k_{il}(T)}{\tilde{\rho}_i}{\tilde{\rho}_l}}
  + \sum_{i} I_i \tilde{\rho}_i
  \label{eqn:hydrospecies}
\end{equation}
for each of the species,
and including the effects of non-equilibrium radiative cooling
and consistent coupling to the hydrodynamics equations.
The ${k_{il}(T)}$ are rate coefficients for the two body reactions
and are tabulated functions of the gas temperature $T$.
The $I_i$ are integrals evaluating the 
photoionization and photodissociation of the different species.
For a comprehensive discussion of the cosmologically 
important chemical reactions and reaction rates, 
see reference~\cite{AAZN97}.

Many numerical techniques have been developed to
solve the hydrodynamic and collisionless particle equations.
For the hydrodynamic equations, the methods range from
Lagrangian SPH algorithms~\cite{Evrard88, HK89, Owen97} to
Eulerian finite difference 
techniques on static meshes~\cite{ROKC93, QIS96}, 
nested grids \cite{ANC94}, moving
meshes~\cite{Gnedin95}, and adaptive mesh refinement~\cite{BN99}.
For the dark matter equations, the canonical choices are
treecodes~\cite{WS95} or
PM and P$^3$M methods~\cite{HE88, EDFW85}, although many variants
have been developed to optimize computational performance
and accuracy,
including grid and particle refinement methods 
(see references cited in~\cite{Frenk99}).
An efficient method for solving non-equilibrium, multi-species
chemical reactive flows together with the hydrodynamic equations
in a background FLRW model is described in~\cite{AAZN97, AZAN97}.

The reader is referred to~\cite{KOCRHEBN94, Frenk99}
for thorough comparisons of different numerical methods applied
to problems of structure formation. Reference~\cite{KOCRHEBN94}
compares (by binning data at different
resolutions) the statistical
performance of five codes (three Eulerian and two SPH)
on the problem of an evolving CDM Universe on large scales
using the same initial data. The results indicate that global
averages of physical attributes converge in rebinned
data, but that some uncertainties remain at small levels.
\cite{Frenk99} compares 
twelve Lagrangian and Eulerian hydrodynamics
codes to resolve the formation of a single X-ray cluster in a CDM
Universe. The study finds generally good agreement for both
dynamical and thermodynamical quantities, but also shows
significant differences in the X-ray luminosity, a quantity
that is especially sensitive to resolution~\cite{AN96a}.


\subsubsection{Linear initial data}
\label{sec:lineardata}

The standard Zel'dovich solution~\cite{Zeldovich70, EDFW85}
can be used to generate initial conditions satisfying
observed or theoretical power spectra of matter density
fluctuations. Comoving physical displacements and
velocities of the collisionless dark matter particles
are set according to the power spectrum realization
\begin{equation}
  \left|\frac{\delta\rho}{\rho}(k)\right|^2\propto k^n T^2(k),
  \label{powerspec}
\end{equation}
where the complex phases are chosen from a gaussian random field,
$T(k)$ is a transfer function~\cite{BBKS86}
appropriate to a particular
structure formation scenario (e.g., CDM), and
$n=1$ corresponds to the Harrison-Zel'dovich power spectrum.
The fluctuations are normalized with top hat smoothing using
\begin{equation}
  \sigma_8^2=\frac{1}{b^2}=\int_0^{\infty} 4\pi k^2 P(k) W^2(k) dk,
  \label{sigma}
\end{equation}
where $b$ is the bias factor chosen to match present observations
of rms density fluctuations in a spherical window of radius
$R_{\rm h} = 8 \, h^{-1}$~Mpc. Also,
$P(k)$ is the Fourier transform of the 
square of the density fluctuations in equation~(\ref{powerspec}), and
\begin{equation}
  W(k)=\frac{3}{(kR_{\rm h})^3}\left(\sin(kR_{\rm h}) -
  (kR_{\rm h})\cos(kR_{\rm h})\right)
\end{equation}
is the Fourier transform of a spherical window of radius $R_{\rm h}$.

Overdensity peaks can be filtered on specified spatial or
mass scales by Gaussian smoothing the random density
field~\cite{BBKS86}
\begin{equation}
  \sigma(r_{\rm o})=\frac{1}{(2\pi R_{\rm f}^2)^{3/2}}\int
  \frac{\delta\rho}{\rho}(r') \exp{\left( 
  -\frac{|r_{\rm o}-r'|^2}{2R_{\rm f}^2}\right)} d^3r'
\end{equation}
on a comoving scale $R_{\rm f}$ centered at $r=r_{\rm o}$
(for example, $R_{\rm f}=5 \, h^{-1}$~Mpc with a filtered mass of
$M_{\rm f} \sim 10^{15} M_\odot$ over cluster scales).
$N\sigma$ peaks are generated by
sampling different random field realizations to satisfy the condition
$\nu=\sigma(r_{\rm o})/\sigma(R_{\rm f})=N$,
where $\sigma(R_{\rm f})$ is the rms of Gaussian filtered density
fluctuations over a spherical volume of radius $R_{\rm f}$.

Bertschinger~\cite{Bertschinger95} has provided a useful 
and publicly available package of
programs called COSMICS for computing transfer functions, CMB
anisotropies, and gaussian random initial conditions for
numerical structure formation calculations. The package solves
the coupled linearized Einstein, Boltzman, and fluid equations for
scalar metric perturbations, photons, neutrinos, baryons,
and collisionless dark matter in a background isotropic Universe.
It also generates constrained or unconstrained matter distributions
over arbitrarily specifiable spatial or mass scales.

\newpage


\bibliography{cosmology-lr}

\end{document}